\RequirePackage{fix-cm}

\documentclass[twocolumn,epjc3]{svjour3}  
\smartqed  
\RequirePackage{mathptmx}      
\RequirePackage{graphicx}
\RequirePackage{amssymb}
\RequirePackage{amsmath}
\RequirePackage{lineno}
\RequirePackage{booktabs}
\RequirePackage{float}
\RequirePackage{comment}
\RequirePackage{mhchem}
\RequirePackage{booktabs}
\RequirePackage{url}
\RequirePackage[colorlinks,allcolors=red]{hyperref}
\RequirePackage[multiple]{footmisc}
\RequirePackage{upgreek}
\RequirePackage{mathtools}
\RequirePackage{cuted}
\RequirePackage{relsize}
\RequirePackage{makecell}
\RequirePackage{tablefootnote}
\RequirePackage{siunitx}

\newcommand{\RomanNumeralCaps}[1]
    {\MakeUppercase{\romannumeral #1}}

\journalname{Eur. Phys. J. C}

\newcommand{\TRIUMF}{TRIUMF, Vancouver, British Columbia V6T 2A3, Canada}
\newcommand{\IHEP}{Institute of High Energy Physics, Chinese Academy of Sciences, Beijing 100049, China}
\newcommand{\BNL}{Brookhaven National Laboratory, Upton, NY 11973, USA}
\newcommand{\McGill}{Physics Department, McGill University, Montr{\'e}al, Qu{\'e}bec H3A 2T8, Canada}
\newcommand{\UMass}{Amherst Center for Fundamental Interactions and Physics Department, University of Massachusetts, Amherst, MA 01003, USA}
\newcommand{\Yale}{Wright Laboratory, Department of Physics, Yale University, New Haven, CT 06511, USA}
\newcommand{\UBC}{Department of Physics and Astronomy, University of British Columbia, Vancouver, British Columbia V6T 1Z1, Canada}
\newcommand{\Stanford}{Physics Department, Stanford University, Stanford, California 94305, USA}
\newcommand{\Erlangen}{Erlangen Centre for Astroparticle Physics (ECAP), Friedrich-Alexander University Erlangen-N\"urnberg, Erlangen 91058, Germany}
\newcommand{\PNL}{Pacific Northwest National Laboratory, Richland, WA 99352, USA}
\newcommand{\Carleton}{Department of Physics, Carleton University, Ottawa, Ontario K1S 5B6, Canada}
\newcommand{\Illinois}{Physics Department, University of Illinois, Urbana-Champaign, IL 61801, USA}
\newcommand{\ITEP}{National Research Center ``Kurchatov Institute", Moscow 117218, Russia}
\newcommand{\SD}{Department of Physics, University of South Dakota, Vermillion, SD 57069, USA}
\newcommand{\LLNL}{Lawrence Livermore National Laboratory, Livermore, CA 94550, USA}
\newcommand{\RPI}{Department of Physics, Applied Physics and Astronomy, Rensselaer Polytechnic Institute, Troy, NY 12180, USA}
\newcommand{\IME}{Institute of Microelectronics, Chinese Academy of Sciences, Beijing 100029, China}
\newcommand{\Colorado}{Physics Department, Colorado State University, Fort Collins, CO 80523, USA}
\newcommand{\Sherbrooke}{Universit\'e de Sherbrooke, Sherbrooke, Qu{\'e}bec J1K 2R1, Canada}
\newcommand{\Laurentian}{School of Natural Sciences, Laurentian University, Sudbury, Ontario P3E 2C6, Canada}
\newcommand{\Wilmington}{Department of Physics and Physical Oceanography, University of North Carolina at Wilmington, Wilmington, NC 28403, USA}
\newcommand{\Indiana}{Department of Physics and CEEM, Indiana University, Bloomington, IN 47405, USA}
\newcommand{\Drexel}{Department of Physics, Drexel University, Philadelphia, PA 19104, USA}
\newcommand{\SLAC}{SLAC National Accelerator Laboratory, Menlo Park, CA 94025, USA}
\newcommand{\ORNL}{Oak Ridge National Laboratory, Oak Ridge, TN 37831, USA}
\newcommand{\Bama}{Department of Physics and Astronomy, University of Alabama, Tuscaloosa, AL 35487, USA}
\newcommand{\IBS}{IBS Center for Underground Physics, Daejeon 34126, Korea}

\newcommand{\UCSD}{Department of Physics, University of California San Diego, La Jolla, CA 92093, USA}
\newcommand{\UWC}{Department of Physics and Astronomy, University of the Western Cape, P/B X17 Bellville 7535, South Africa}
\newcommand{\SNOLAB}{SNOLAB, Sudbury, ON P3E 2C6, Canada}
\newcommand{\Kentucky}{Department of Physics and Astronomy, University of Kentucky, Lexington, KY 40506}
\newcommand{\Skyline}{Skyline College, Skyline College, San Bruno, CA 94066, USA}
\newcommand{\mines}{Colorado school of mines, Department of Physics, Colorado School of Mines, Golden, CO 80401, USA}
\newcommand{\SUBATECH}{SUBATECH, IMT Atlantique, CNRS/IN2P3, Universit\'e de Nantes, Nantes 44307, France}

\begin{document}

\title{Performance of novel VUV-sensitive Silicon Photo-Multipliers for nEXO}
 \widowpenalty10000
  \clubpenalty10000

\institute{
            \TRIUMF\label{l_TRIUMF}
            \and
            \IHEP \label{l_IHEP}
            \and
            \BNL \label{l_BNL}
             \and
             \Drexel \label{l_Drexel}
            \and
            \McGill \label{l_McGill}
            \and
             \UMass \label{l_UMass}
             \and
            \Yale \label{l_Yale}
             \and
            \UCSD\label{l_UCSD}
              \and
           \Stanford \label{l_Stanford}
            \and
           \PNL \label{l_PNL}
            \and
            \Sherbrooke \label{l_Sherbrooke}
           \and
           \Carleton \label{l_Carleton}
            \and
            \ITEP \label{l_ITEP}
            \and
            \LLNL \label{l_LLNL}
             \and
            \Kentucky\label{l_Kentucky}
            \and
            \SLAC \label{l_SLAC}
             \and
            \RPI \label{l_RPI}
            \and
             \Laurentian \label{l_Laurentian}
            \and
            \SNOLAB\label{l_SNOLAB}
            \and
           \IME \label{l_IME}
            \and
            \Bama \label{l_Bama}
            \and
           \Wilmington \label{l_Wilmington}
            \and
           \Illinois \label{l_Illinois}
            \and
           \ORNL \label{l_ORNL}
                      \and
           \Colorado \label{l_Colorado}
                       \and
             \Skyline\label{l_Skyline}
                          \and
           \Erlangen \label{l_Erlangen}
           \and
           \UBC \label{l_UBC}
             \and
           \SD \label{l_SD}
            \and
             \mines\label{l_mines}
             \and
           \IBS \label{l_IBS}
            \and
            \UWC \label{l_UWC}
            \and
             \SUBATECH\label{l_SUBATECH}
            \and
           \Indiana \label{l_Indiana}
}

\author{G.~Gallina\thanksref{l_TRIUMF,corr1,l_PRINCETON}
        \and
        Y.~Guan\thanksref{l_IHEP,l_CAS}
        \and
        F.~Retiere\thanksref{l_TRIUMF}
        \and
        G.~Cao\thanksref{l_IHEP,l_CAS,corr2}
        \and
        A.~Bolotnikov\thanksref{l_BNL}
        \and
        I.~Kotov\thanksref{l_BNL}
         \and
         S.~Rescia\thanksref{l_BNL}
          \and
         A.K.~Soma\thanksref{l_Drexel}  
          \and
         T.~Tsang\thanksref{l_BNL}
        \and
         L.~Darroch\thanksref{l_McGill}
% ...
}

\thankstext{corr1}{e-mail: giacomo@triumf.ca (corresponding author)}
\thankstext{l_PRINCETON}{Also with: Physics Department, Princeton University, Princeton, NJ 08544, USA}
\thankstext{l_CAS}{Also with: University of Chinese Academy of Sciences, Beijing 100049, China}
\thankstext{corr2}{e-mail: caogf@ihep.ac.cn (corresponding author)}
\thankstext{l_MIT}{Now at: Laboratory for Nuclear Science, Massachusetts Institute of Technology, Cambridge, MA, USA}
\thankstext{l_SaudiArabia}{Permanent at: King Abdulaziz City for Science and Technology, KACST, Riyadh 11442, Saudi Arabia}
\thankstext{l_Canon}{Now at: Canon Medical Research, USA}
\thankstext{l_Berkley}{Now at: Nuclear Science Division, Lawrence Berkeley National Laboratory, USA}
\thankstext{l_Zurich}{Now at: Institute for Particle Physics and Astrophysics, ETH Zürich, Switzerland}

\date{Received: date / Accepted: date}

\begingroup
\fontsize{6.5pt}{7.5pt}\selectfont
\maketitle
\endgroup

\modulolinenumbers[5]

\begin{abstract}
    
Liquid xenon time projection chambers are promising detectors to search for neutrinoless double beta decay (0$\nu \beta \beta$), due to their response uniformity, monolithic sensitive volume, scalability to large target masses, and suitability for extremely low background operations. The nEXO collaboration has designed a tonne-scale time projection chamber that aims to search for 0$\nu \beta \beta$ of \ce{^{136}Xe} with projected half-life sensitivity of $1.35\times 10^{28}$~yr. To reach this sensitivity, the design goal for nEXO is $\leq$1\%  energy resolution at the decay $Q$-value ($2458.07\pm 0.31$~keV). Reaching this resolution requires the efficient collection of both the ionization and scintillation produced in the detector. The nEXO design employs Silicon Photo-Multipliers (SiPMs) to detect the vacuum ultra-violet, 175 nm scintillation light of liquid xenon. This paper reports on the characterization of the newest vacuum ultra-violet sensitive Fondazione Bruno Kessler VUVHD3 SiPMs specifically designed for nEXO, as well as new measurements on new test samples of previously characterised Hamamatsu VUV4 Multi Pixel Photon Counters (MPPCs). Various SiPM and MPPC parameters, such as dark noise, gain, direct crosstalk, correlated avalanches and photon detection efficiency were measured as a function of the applied over voltage and wavelength at liquid xenon temperature (163~K).  The results from this study are used to provide updated estimates of the achievable energy resolution at the decay $Q$-value for the nEXO design.  
\end{abstract}

\section{Introduction}
\label{S:intro}

Silicon Photo-Multiplier (SiPM) represents an excellent solid-state photon detection technology, combining the low-light detection capabilities of conventional vacuum photo-multiplier tubes (PMTs) with the benefits of solid-state sensors. In contrast to PMTs or to large-area avalanche photodiodes, SiPMs consist of an array of tightly packaged Single Photon Avalanche Diodes (SPADs) with quenching resistors operated above the breakdown voltage, $V_{bd}$, to generate self-sustaining charge avalanches upon absorbing incident photons \cite{gallina2021development}.\\\indent Generally, SiPMs are 
a compelling photosensor solution  when operated in liquid noble gases due to their very low residual natural radioactivity, low-voltage operation, compact and flat form factor~\cite{Baudis2018}. For these reasons, SiPMs are the baseline solution in the MEG\RomanNumeralCaps{2} experiment \cite{Baldini2018}, that is currently collecting physics data in Liquid Xenon (LXe), and in the planned DUNE \cite{FALCONE2021164648} and DarkSide-20k experiments  \cite{aalseth2018darkside,SiPMFBKDS20k}. The nEXO detector is a planned double beta decay experiment that aims to probe the boundaries of the standard model of particle physics by searching for 0$\nu \beta \beta$ of \ce{^{136}Xe}. Its projected half-life sensitivity after 10 years of data taking is expected to be $1.35\times 10^{28}$~yr at 90\% confidence level \cite{Adhikari2022} with a final design goal of $\leq$1\%  energy resolution at the decay $Q$-value ($2458.07\pm 0.31$~keV \cite{Redshaw2007,PhysRevC.82.024603}\footnote{Weighted average of the values published in Refs.~\cite{Redshaw2007,PhysRevC.82.024603}.}). The nEXO detector follows the concept of the previous generation EXO-200 detector and uses five tonnes of LXe, enriched to \ce{^{136}Xe}, inside a Time Projection Chamber (TPC) with both charge and scintillation light readout. It is planned to be operated at SNOLAB, the Canadian underground science laboratory~\cite{NEXOCollaboration2018,Conti2003}.\\\indent Large LXe detectors for dark matter searches typically employ PMTs as light detectors. However, even the low-background PMTs developed for use in such detectors~\cite{Abe2019,Aprile2015} have U/Th content that exceeds the nEXO background requirements~\cite{Adhikari2022} by more than a factor of 1000x, precluding their consideration.\\\indent Vacuum Ultra-Violet (VUV) sensitive detectors fabricated from silicon ({\it i.e.} avalanche photodiodes (APDs) and SiPMs) have substantially lower backgrounds~\cite{Baudis2018}. However, among silicon detectors, SiPMs are expected to have the lowest backgrounds, in addition, of being low-voltage powered and optimal for operation at cryogenic temperatures. Moreover, there is a significant risk in identifying a commercial supplier for large area APDs, such as the ones used in EXO-200~\cite{Neilson2009}, since they are no longer commercially available in the market. For these reasons, the nEXO collaboration selected SiPMs as the photosensor of choice to detect the VUV scintillation light of LXe (mean wavelength of $174.8$~nm and full width at half maximum (FWHM) of $10.2$~nm \cite{FUJII2015293}). As a result of recent developments by a variety of photonics companies and research institutions, SiPMs now have wavelength sensitivity extending into the 175~nm region. In 2016, Fondazione Bruno Kessler (FBK) produced several batches of VUV-sensitive SiPMs for nEXO (FBK VUVHD1) \cite{Ako}. Hamamatsu Photonics Inc. (HPK) also developed new generations  of VUV sensitive MPPCs (HPK VUV4) for applications in LXe~\cite{Baldini2018,Ieki2019LargeareaMW,Baudis_2021}.\\\indent The photon detection system of the nEXO experiment must be consistent with nEXO's challenging background goals~\cite{Adhikari2022}, and must meet the following requirements (in LXe, whose boiling temperature is $165.02$~K \cite{Brown1980}): (i) Photon Detection Efficiency (PDE) greater than 15\% for $175$~nm photons, (ii) Dark count rate $<10~\text{Hz}/\text{mm}^2$, (iii) Operational gain larger than $1.5\times10^6$ electrons per photo-electron, (iv) Correlated avalanche fluctuation per pulse in the $1\upmu\text{s}$ window following the trigger pulse $<0.4$\footnote{In nEXO, the expected charge integration time after the trigger pulse will be up to 1$\upmu\text{s}$ long.}. This last quantity is defined as the ratio between the root mean square error and the average extra charge produced by correlated avalanches per pulse and essentially set an upper limit on the SiPM correlated noise on an avalanche-to-avalanche basis, more details are reported in Sec.~\ref{S:CA}. In addition to the four requirements on the SiPMs photosensors, the electronic noise of the nEXO photon detection system needs to be smaller than 0.1 Photo-electron Equivalent (PE) r.m.s.\footnote{The last requirement is a combination of power consumption constraint and the total area that one channel of the front-end electronics can read out \cite{fabristhesis}.}\cite{Adhikari2022}. The above requirements differ slightly from the ones previously described in Ref.~\cite{NEXOCollaboration2018} due to an improved understanding of the detector energy resolution model that will be discussed in Sec.~\ref{S:energyresolution}. The first generation of FBK SiPMs (FBK VUVHD1) comfortably met the nEXO PDE requirement \cite{Ako}, while the HPK VUV4 MPPCs previously tested only marginally met it~\cite{Gallina2019}. 
\\\indent The aim of this work is to assess the performance of the newest generation of FBK devices specifically designed for the nEXO experiment (FBK~VUVHD3) and to present the results from new measurements of two types of commercially available VUV sensitive  HPK VUV4 MPCCs: HPK 
S13370-6050CN (HPK VUV4-50) and HPK S13371-6050CQ (HPK VUV4-Q-50), as shown in Table~\ref{tab:SiPM_values} (abbreviations used in plots are in parentheses). Compared to the previous generation of FBK devices, the primary change introduced in the FBK~VUVHD3 SiPMs is a novel triple-doping technology that suppresses SiPM afterpulses~\cite{Capasso2020} and therefore increases their operational over voltage while still satisfying other nEXO requirements.  Differently from the HPK MPPCs that were  purchased independently by several nEXO institutions and  therefore might be from different production batches, the FBK VUVHD3 SiPMs  tested in this work are extracted from the same engineering wafer produced by FBK for nEXO.\\\indent Several SiPM characteristics such as gain, dark count rate and correlated avalanche fluctuation that can potentially impact the nEXO energy resolution are studied at $163$~K as a function of the applied over voltage in dedicated test setups and compared against the nEXO requirements. Overall 11 FBK VUVHD3 SiPMs and 6 HPK MPPCs (4 HPK VUV4-Q-50 MPPCs, 2 HPK VUV4-50 MPPCs) were characterized in this study at test setups at six institutions.  Each setup operated SiPMs pursuing different objectives. Thus, not all properties were measured by each setup and for every tested device.   Specifically, each photosensor noise characteristic, correlated or not, was measured at least on two  photosensors at two or more institutions. The PDE of 2 FBK VUVHD3 SiPMs and 3 HPK MPPCs (2 HPK VUV4-Q-50 MPPCs and 1 HPK VUV4-50 MPPC) was also measured as a function of the excess voltage above breakdown called over voltage {\it i.e.} $(V - V_{bd})$ with $V$ reverse bias voltage, both at 300~K and 163~K in the (165--200)~nm wavelength range (LXe scintillation emission spectrum is in the shape of a Gaussian function, with the mean at $174.8~\text{nm}$ and a FWHM of  $10.2~\text{nm}$ \cite{FUJII2015293}). The variety of devices, test conditions and setups provides good grounds to consider this a representative sample of the behaviour of the SiPMs under consideration for nEXO. These results are then used to infer the performance of the nEXO detector in terms of the achievable energy resolution.
\begin{table*} [ht]
\centering
\begin{tabular}{ccccc}
   \toprule
    & FBK VUVHD1 &  FBK VUVHD3 & \multicolumn{2}{c}{HPK VUV4 MPPCs}  \\
   \midrule
   S/N & - & - & S13370-6050CN & S13371-6050CQ\\
   PA [$\text{mm}^2$] & 5.96$\times$5.56 & 5.96$\times$5.56 & 6$\times$6 & $4\times$(5.95$\times$5.85) \\
    Abbreviation & VUVHD1 &  VUVHD3 & VUV4-50 & VUV4-Q-50\\
Pitch [$\upmu\text{m}^2$] & $35\times35$ & $35\times35$ & $50\times50$ & $50\times50$\\
 Window & \multicolumn{4}{c}{Bare (unsealed)}\\
   \bottomrule
\end{tabular} 
\caption{Summary of the physical properties of the FBK VUVHD3 SiPMs and HPK MPPCs characterised in this work and abbreviations used in the text to identify them. All the FBK SiPMs tested in this work are extracted from the same wafer produced by FBK for nEXO. (PA) stands for photosensitive area. Serial Number (S/N) of HPK MPPCs refers to the documentation published in Ref.~\cite{Hamamatsu_doc}. The -Q- in the abbreviation of the HPK VUV4 (S13371-6050CQ) highlights the fact that these MPPCs are quad (4) devices mounted on the same ceramic package.  In the table, we also report the characteristics of the previous generation of FBK SiPM (FBK VUVHD1) tested in Ref.~\cite{Ako} and of the HPK VUV4 (S13371-6050CN), previously characterised in Ref.~\cite{Gallina2019}.}
 \label{tab:SiPM_values}
\end{table*}

\section{Hardware Setups}
\label{S:hardware}

Several cryogenic test setups were developed within the nEXO collaboration to characterize the SiPMs at VUV wavelengths. Table~\ref{tab:harware} summarizes the main hardware components and the data acquisition (DAQ) systems of those used for this work. Specifically, we present new results from seven test setups, distributed as follows among nEXO institutions (abbreviations used in plots are in parentheses):
two in Canada, at TRIUMF (TR) and McGill University (MG); three in the USA, at Yale University (YALE), the University of Massachusetts, Amherst (UMASS), and Brookhaven National Laboratory (BNL); two in China, both at the Institute of High Energy Physics (IHEP). Results previously obtained with three more setups, two in the USA, at Stanford University (ST) and the University of Alabama (AL), and one in Germany, at the Erlangen Center for Particle Astrophysics (ER), and described elsewhere~\cite{Ako,Nakarmi2020,Wagenpfeil2021}, are also included in this work, for comparison. 
\\\indent In general, the measurements presented in this paper were made in vacuum conditions using dry cryostats in which the SiPM devices are mounted on supports coupled to cold fingers that set and maintain their temperature down to 163~K, with temperature stability better than 0.5~K for all setups. While the intrinsic SiPMs characteristics ({\it e.g.} dark and correlated avalanche noise) are expected to be mostly unchanged while operating them in LXe, the SiPMs PDE could slightly differ due to the change of the medium index of refraction. The difference in the detection medium is therefore accounted for in the nEXO simulation framework by combining the vacuum-measured SiPM PDE data, presented in this work, with the corresponding SiPM reflectivity, measured both in vacuum and LXe, previously reported in Refs.~\cite{Nakarmi2020,Lv2020,Wagenpfeil2021}. More details on the nEXO simulations can be found in Ref.~\cite{nEXO_Sensitivity}.\\\indent For pulse counting measurements, the SiPM signal was either amplified with fast wideband RF amplifiers (TRIUMF, BNL, McGill), or with Cremat charge sensitive preamplifiers  and Gaussian shapers (other groups). The two approaches are mostly equivalent other than for a partial loss in the resolution of overlapping pulses with the latter approach. CAEN digitizers or fast oscilloscopes constitute the DAQ systems of the nEXO testing setups. SiPM reverse bias I-V curves and NIST calibrated diode photo-currents used for PDE measurements to calibrate the absolute light flux at the SiPM location in two of the involved setups (TRIUMF and IHEP), were measured with low noise picoammeters. The nEXO test setups at Yale, Umass, Alabama include a LXe purification and liquefaction system, which allow SiPM testing in LXe as well as in vacuum and Xe gas. These test setups, together with larger scale ones operated at Stanford University and McGill University, will be used for future studies of the long-term stability of SiPMs in nEXO-like operating conditions and to test large arrays of SiPMs.

\begingroup
\renewcommand{\arraystretch}{2}
\setlength{\tabcolsep}{5pt}
\begin{table*} [ht]
\centering
\begin{tabular}{ccccccc}
   \toprule
    & TRIUMF &  \makecell{McGill\\University} & \makecell{Yale\\University} & \makecell{University of Mas-\\sachusetts,Amherst} & \makecell{Brookhaven Natio-\\nal Laboratory~\cite{Kotov2016}} & \makecell{Institute of High\\Energy Physics}\\
   \midrule   Abbreviation & TR & MG & YALE & UMASS & BNL & IHEP \\ \makecell{Temperature\\Stabilisation} & Instec MK2000 & Lakeshore 350 & \makecell{custom\\LabVIEW} & \makecell{custom\\LabVIEW} & CryoCon 24C &  \makecell{ CTE-SG12012\\-02W}\\
\makecell{Measurement\\Temperature} & 163~K & 163~K & 163~K & 190~K/163~K & 163~K & 300~K/233~K\\
 \makecell{SiPM\\Amplification}  & \makecell{MAR6-SM+\\OPA695~\cite{Amp}} & \makecell{MAR6-SM+\\OPA695~\cite{Amp}} & \makecell{CR-113-R2\\SRS SR-560} & \makecell{CR-113-R2\\CR-200-100ns} & \makecell{MAR6-SM+\\OPA695~\cite{Amp}} & \makecell{custom\\amplifier \cite{fabristhesis}}\\
 \makecell{DAQ\\pulse counting} &   \makecell{CAEN\\DT5730B} & \makecell{Rohde \& Schwarz\\RTO2024} & \makecell{Rohde \& Schwarz\\RTB2004} & \makecell{Teradyne\\ZTEC ZT4421} & \makecell{MSO64\\Tektronix} &   \makecell{CAEN\\DT5751}\\
 DAQ I-V & \makecell{Keithley 6487\\Keysight B2985A} & Keysight B2987 & Keithley 6487 & Keithley 6482 & - & Keithley 6487\\
 LXe/GXe & No & No & Yes & Yes & No & No\\
 \makecell{SiPM Noise\\analysis} & Yes & Yes & Yes & Yes & Yes & Yes\\
 SiPM PDE & Yes & No & No & No & No & Yes\\
   \bottomrule
\end{tabular} 
\caption{Hardware components and Data Acquisition (DAQ) for pulse counting, reverse bias I-V curves and diode photocurrent measurements (used to calibrate the absolute light flux, Sec.~\ref{S:PDE}) of the cryogenic test setup developed within the nEXO collaboration and used for this work. Specifically, in this work we present new results from seven test setups, distributed among nEXO institutions. Results previously obtained with three more setups: at Stanford University (ST), University of Alabama (AL), and at the Erlangen Center for Particle Astrophysics (ER), and described elsewhere~\cite{Ako,Nakarmi2020,Wagenpfeil2021}, are also included in this work, for comparison. The nEXO test setups at Yale, UMass and Alabama include a Liquid Xenon (LXe) purification and liquefaction system, which allow SiPM testing in LXe as well as in vacuum and gas Xe (GXe). 
}
 \label{tab:harware}
\end{table*}
\endgroup

\subsection{Collected Data and Trigger Configurations}

The SiPM data were collected following the scheme presented in  Ref.~\cite{Gallina2019}, and can be divided in two sub-categories: dark data and continuous lamp data. Dark measurements were made by the nEXO institutions at 163~K and at multiple over voltages with the photosensors operated in dark conditions. For each setup, the DAQ triggered on individual dark pulses with a DAQ threshold above the noise (Sec.~\ref{S:gain}, Sec.~\ref{S:CDP}). Lamp driven measurements, {\it i.e.}, PDE data, were collected as a function of the applied over voltage by TRIUMF and IHEP at temperatures of 163~K and 300~K, respectively. Due to the different temperatures, the two institutions used different techniques for measuring the photosensors PDE. Overall, the two procedures involve the measurement of the photosensors in pulse counting mode, as well as the measurement of their photo-current under illumination. The measurements taking place at TRIUMF had the DAQ triggered by individual lamp driven pulses with a DAQ threshold above noise, while the measurements at IHEP were externally triggered from a waveform generator, which also fired an LED (404~nm) for the evaluation of a correction factor used for the PDE measurement technique. More details can be found in Sec.~\ref{S:PDE}. 

\section{Experimental Results} 
\label{S:ExpRes}
\subsection{Signal Pulse Analysis Procedure} 
\label{S:fitsection}

The dark data were analyzed at the pulse level by each of the involved nEXO institutions. Single PE pulses were either fitted by TRIUMF and Yale using waveform analysis toolkits similar to the one presented in Refs.~\cite{Ako,Gallina2019} to deconvolve overlapping pulses and extract the pulse time and area, or integrated numerically by other institutions. The two approaches are equivalent for what concerns the measurement of the SiPM gain (Sec.~\ref{S:gain}), and of the correlated avalanche fluctuation (Sec.~\ref{S:CA}).
The TRIUMF and Yale identification and fitting follow the same scheme presented in Ref.~\cite{Gallina2019}, and rely on a ${\chi}^{2}$ minimization to identify and fit SiPM pulses. First, a pulse-finding algorithm based on a matched-filter scheme identifies and fits single avalanche pulses to extrapolate the average SiPM pulse shape, parameterised as shown in Refs.~\cite{gallina2021development,Gallina2019}. The SiPM pulse shape is then set by fixing these parameters to their estimated average values. Finally, a second fit iteration is performed with fixed pulse shape to improve the estimation of pulse time and area. For fits exceeding a certain reduced ${\chi}^{2}$ threshold, multiple pulses are added iteratively to the fit. The new pulse combination is adopted if the reduced ${\chi}^{2}$ of the new fit improves significantly. Otherwise, the added pulses are discarded. The last step of the algorithm improves the capability to identify overlapping pulses. More details can be found in Ref.~\cite{Retiere2009}.

\subsection{Single PE Gain and breakdown voltage Extrapolation}
\label{S:gain}

\begin{figure}[ht]
\centering
\includegraphics[width=0.99\linewidth]{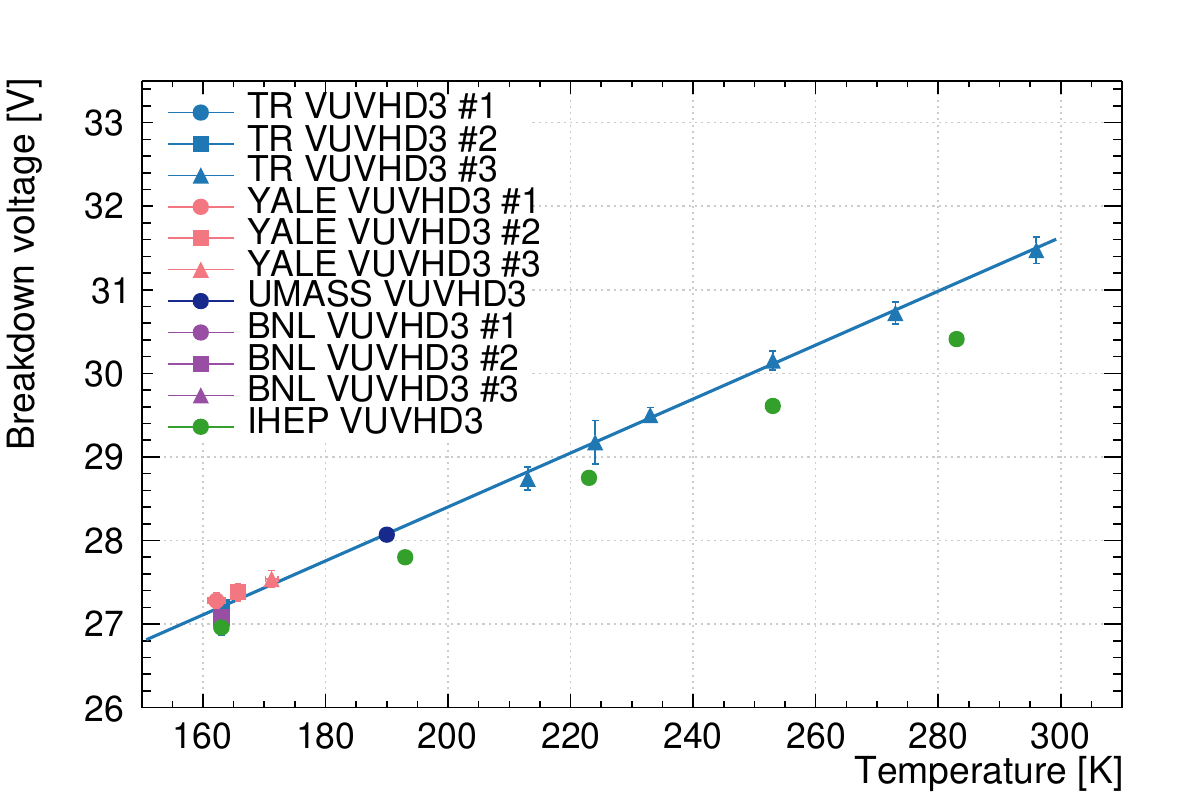}
\includegraphics[width=0.99\linewidth]{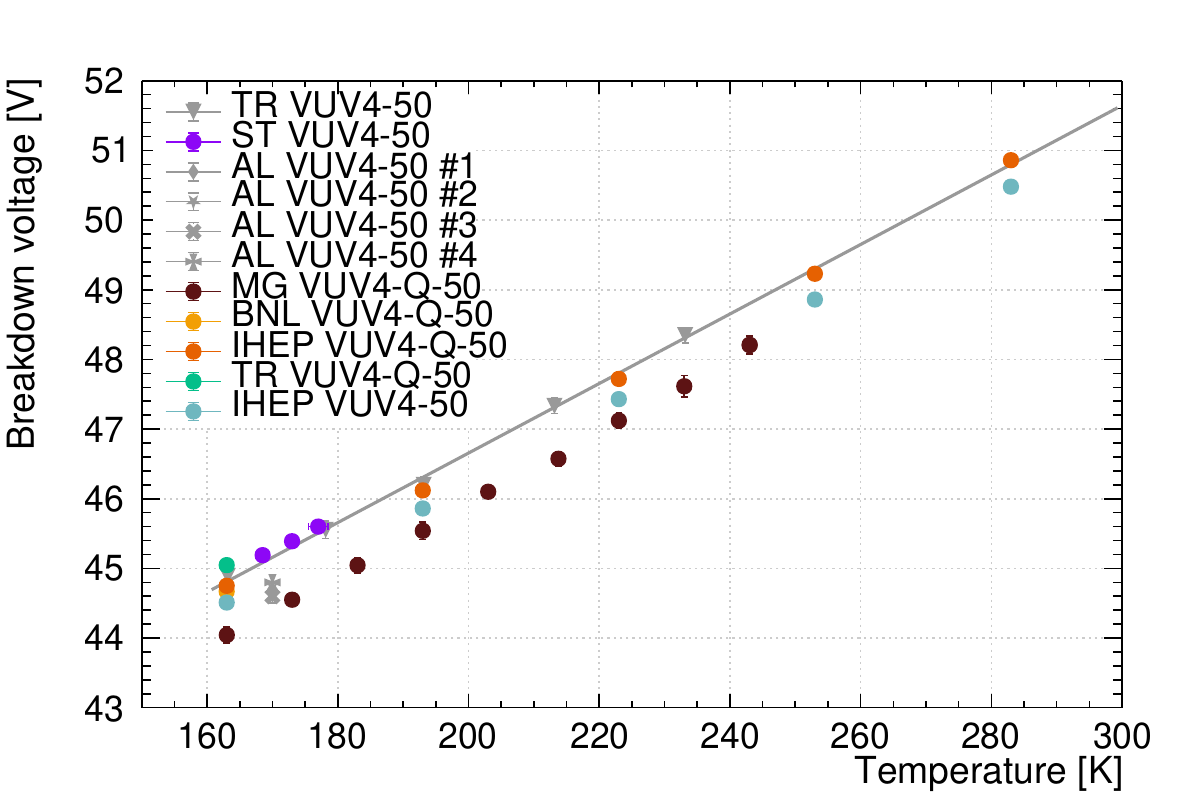}
\caption{Breakdown voltage measured, as a function of the temperature, for FBK VUVHD3 SiPMs (top) and HPK VUV4 MPPCs (bottom). The lines represent fits in order to extract the breakdown temperature gradient. Only the fits of the TRIUMF data are shown for clarity.}
\label{fig:vbd_hd3_vuv4}
\end{figure}

\begin{figure}[ht]
\centering
\includegraphics[width=0.99\linewidth]{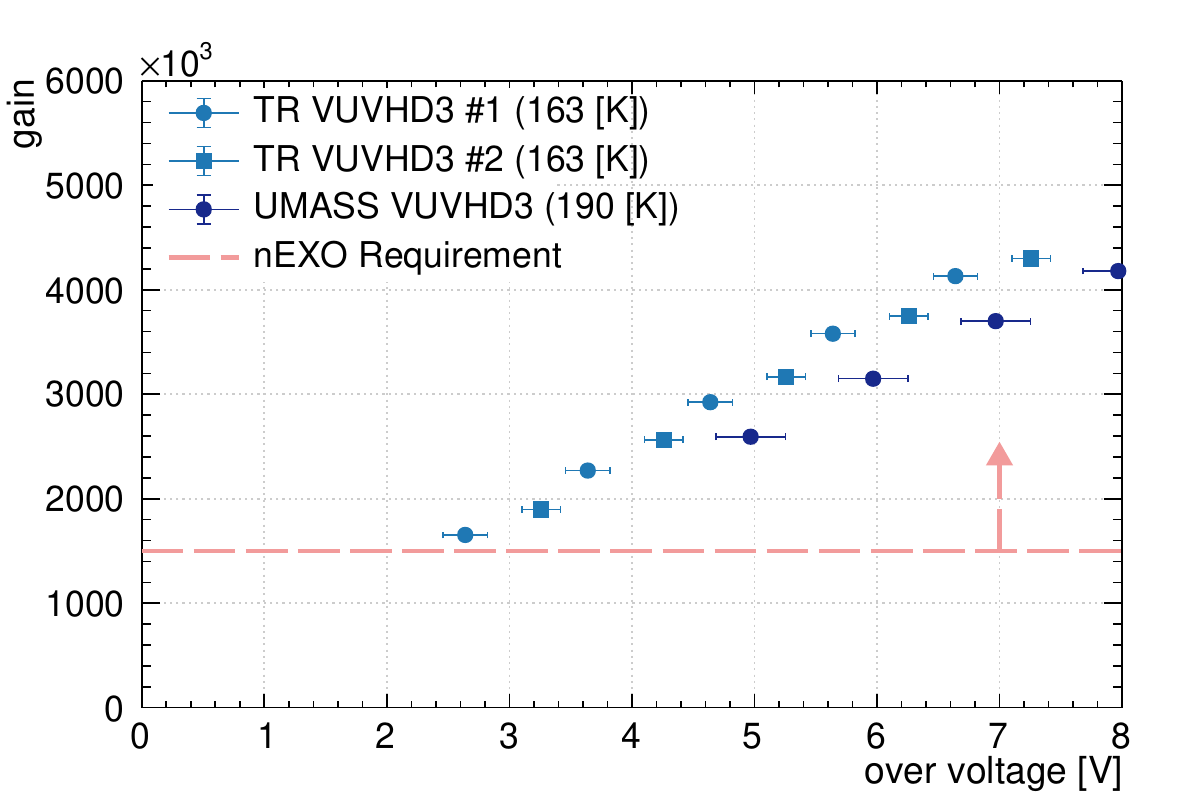}
\includegraphics[width=0.99\linewidth]{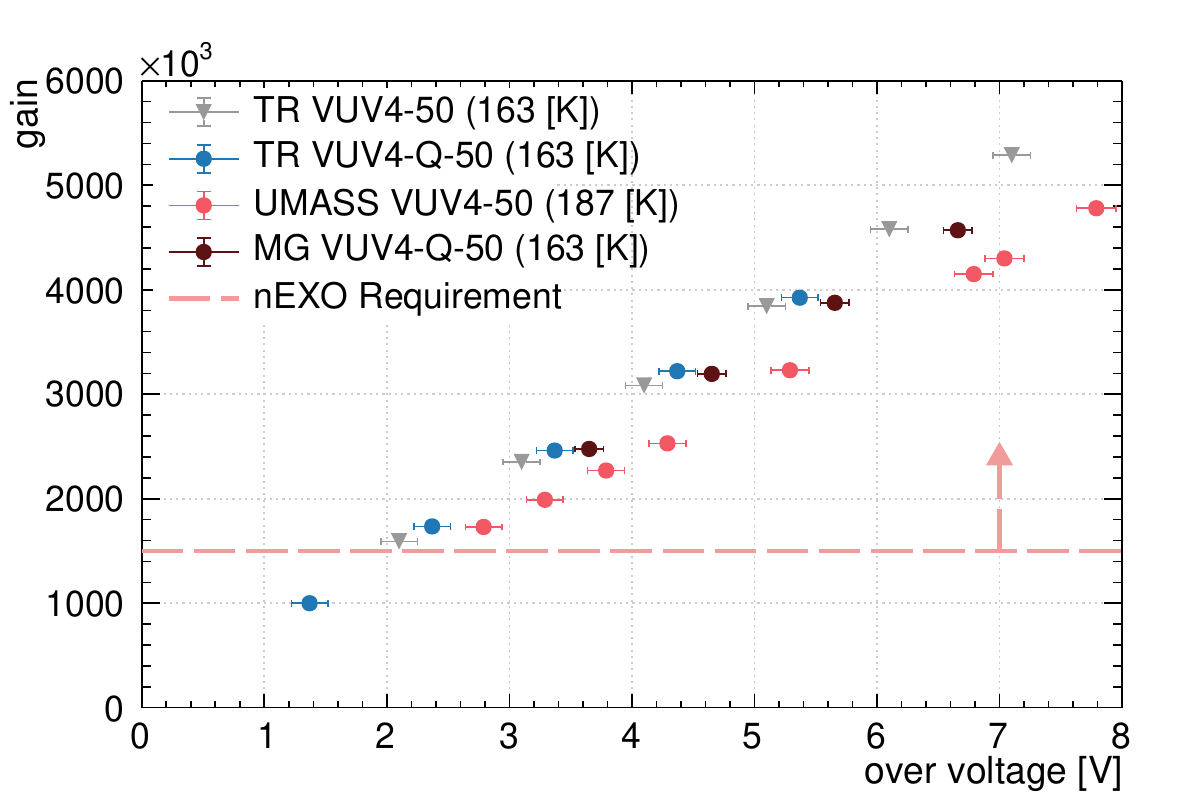}
\caption{Single PE gain (Eq.~\ref{eq:gain}) measured, as a function of the over voltage, for FBK VUVHD3 SiPMs (top) and  HPK VUV4 MPPCs (bottom). The dashed line represents the nEXO requirement.}
\label{F:SPE_gain_FBK_HPK}
\end{figure}

Single PE pulses were used to compute the average single PE charge ($\overline{\text{Q}}_{1\text{ PE}}$) either by fitting or by numerical integration. From the single PE charge, after calibration of the readout electronics\footnote{The DAQ of each institution was separately calibrated by applying a step voltage to a precision capacitor to inject a known charge into the SiPM amplifier input.}, it is possible to extrapolate the single PE gain, defined as:
\begin{linenomath}
\begin{equation}
\label{eq:gain}
\overline{\text{G}}_{1\text{ PE}}= \frac{\overline{\text{Q}}_{1\text{ PE}}}{q_{E}}
\end{equation}
\end{linenomath}
where $q_{E}$ is the electron charge. The average single PE charge was then linearly fitted as a function of the applied bias voltage $V$ as follows 
\begin{linenomath}
\begin{equation}
\label{eq:gain_to_zero}
\overline{\text{Q}}_{1\text{ PE}}= C_{D}\times\left( V - V_{bd}\right)
\end{equation}
\end{linenomath}
This is done in order to extract the SiPM single microcell capacitance ($C_{D}$), and the breakdown voltage ($V_{bd}$), defined as the bias voltage for which the average SiPM single PE charge is zero. The breakdown voltage as a function of temperature is shown in Fig.~\ref{fig:vbd_hd3_vuv4} for FBK VUVHD3 SiPMs and HPK VUV4 MPCCs.
The average single PE gain for the same devices is shown in Fig.~\ref{F:SPE_gain_FBK_HPK}.
Table~\ref{T:Summary_gain} summarises the average breakdown voltage and breakdown temperature gradient
obtained performing respectively: (i) a weighted average of all the data of Fig.~\ref{fig:vbd_hd3_vuv4} with a temperature T such that $|\text{T}-165|\le3$~K, (ii) a weighted average of the slopes of the fits of the data for which the breakdown temperature dependence was actually measured (TRIUMF and IHEP data for FBK VUVHD3 SiPMs; TRIUMF, McGill and IHEP data for HPK MPPCs). In addition, Table~\ref{T:Summary_gain} reports the average SPAD capacitance measured at 163~K. This last quantity was obtained by first fitting each dataset of Fig.~\ref{F:SPE_gain_FBK_HPK} with Eq.~\ref{eq:gain_to_zero}, divided by $q_e$ in order to extrapolate the corresponding slopes, and then taken as the weighted average of all measurements. Errors in Table~\ref{T:Summary_gain} are computed by adding the standard error and variance in quadrature.\\\indent The larger SPAD capacitance and gain of the HPK~VUV4 MPPCs is a result of its larger single cell pitch, as shown in Table~\ref{tab:SiPM_values}. Both FBK SiPMs and HPK MPPCs comfortably satisfies the nEXO requirement, represented with an horizontal line in Fig.~\ref{F:SPE_gain_FBK_HPK}, from roughly 2.5~V and 2~V of over voltage, respectively\footnote{From an electronic point of view, we desire to have sufficient Signal to Noise Ratio (SNR) that the electronics noise is negligible (i.e. large gain, large over voltage). From the opposite perspective, the SiPM’s correlated avalanches, studied in Sec.~\ref{S:CA} increase quickly with the applied over voltage that should therefore be limited. The requirement on the SiPM gain essentially set a lower limit on the capability to operate in single photon counting regime, which can be comfortably achieved with an SNR equal to 10, {\it i.e.} a noise level of 0.1 PE r.m.s., as reported in Sec.~\ref{S:intro}.}. Even without any pre-selection of the FBK or HPK devices to test in the various nEXO test setups, we see a remarkably small spread in their breakdown voltages ($\Delta V_{bd}<0.5$~V) at $163~\text{K}$. A relatively larger spread, based on a smaller statistical basis, is seen in the corresponding single PE gain, as shown in Fig.~\ref{F:SPE_gain_FBK_HPK}.

\begin{table}[ht]
\centering
\begin{tabular}{cccc}
   \toprule
   Device & $C_D$ [fF] & $V_{bd}$ [V] & $\Delta V_{bd}/\Delta T$ [mV/K]\\
   \midrule
   HPK~VUV4-50 & $101\pm6$  & $44.51\pm0.05$ & $50\pm2$\\ 
    HPK~VUV4-Q-50 & $111\pm4$  & $44.73\pm0.09$ & $52\pm2$\\
   FBK~VUVHD3 & $90\pm5$  & $27.09\pm0.17$  & $29.1\pm0.9$ \\
   \bottomrule
\end{tabular} 
\caption{Summary of the measured: (i) 163~K  single cell capacitance ($C_D$), (ii) 163~K  average breakdown voltage ($V_{bd}$) and (iii) breakdown temperature gradient ($\Delta V_{bd}/\Delta T$) for all the devices tested in this work. Errors are computed by adding the standard error and variance in quadrature. See text for more details.}
\label{T:Summary_gain}
\end{table}

\subsection{Noise Analyses}
\label{S:CDP}

Dark and correlated avalanche noise are crucial SiPM parameters that can affect the overall nEXO energy resolution by artificially increasing the fluctuations in the number of photons detected by the SiPMs. Dark noise pulses are spontaneous charge signals generated by electron-hole pairs formed by thermal or field enhanced processes \cite{Hurkx1992}. 
nEXO requires a dark count rate $\le~10~\text{ Hz}/\text{mm}^2$, a value mainly driven by the goal to identify low energy scintillation pulses for background rejection.\\\indent Correlated Avalanche (CA) noise is due to at least two processes: the production of secondary photons in the gain amplification stage during primary avalanches and the trapping and subsequent release of charge carriers produced in avalanches (afterpulsing). Afterpulse events trigger the same cell multiple times in which the original avalanche happened. Secondary photons in SiPMs are responsible, instead, for at least three processes: (i) internal crosstalk (ii) external crosstalk and (iii) optically-induced afterpulsing.\\\indent Internal crosstalk refers to the secondary photons that trigger avalanches in neighbouring SPADs of the same SiPM without escaping from the SiPM itself. External crosstalk refers to the secondary photons that escape from the surface of one SPAD and either (i) reflect back into the SiPM at the surface coating interface and trigger avalanches in neighbouring SPADs \cite{gundacker2020silicon}, or (ii) transmit through the SiPM surface coating and leave the SiPM hitting another SiPM \cite{McLaughlin2021}. Finally, optically-induced afterpulsing refers to
the secondary photons that trigger avalanches in the same SPAD where secondary photon emission occurs during the SPAD recharging time. Avalanches inside the same SiPM triggered by secondary photons can be simultaneous with the primary one (Direct CrossTalk (DiCT)) or delayed by several nanoseconds (Delayed CrossTalk (DeCT)) \cite{Boone2017}.\\\indent In general, the subset of the CAs consisting of afterpulses, optically induced or not, and DeCT events is referred to as Correlated Delayed Avalanches (CDAs). The number of CDAs, as well as the SiPMs dark count rate are discussed in Sec.~\ref{S:DN}. Unlike dark noise events, CAs (and therefore CDAs) are correlated with a primary signal and are thus present only if an avalanche happens, {\it i.e.}, a SPAD is discharged with the subsequent production of a pulse.
\begin{figure}[ht]
\centering
\includegraphics[width=0.99\linewidth]{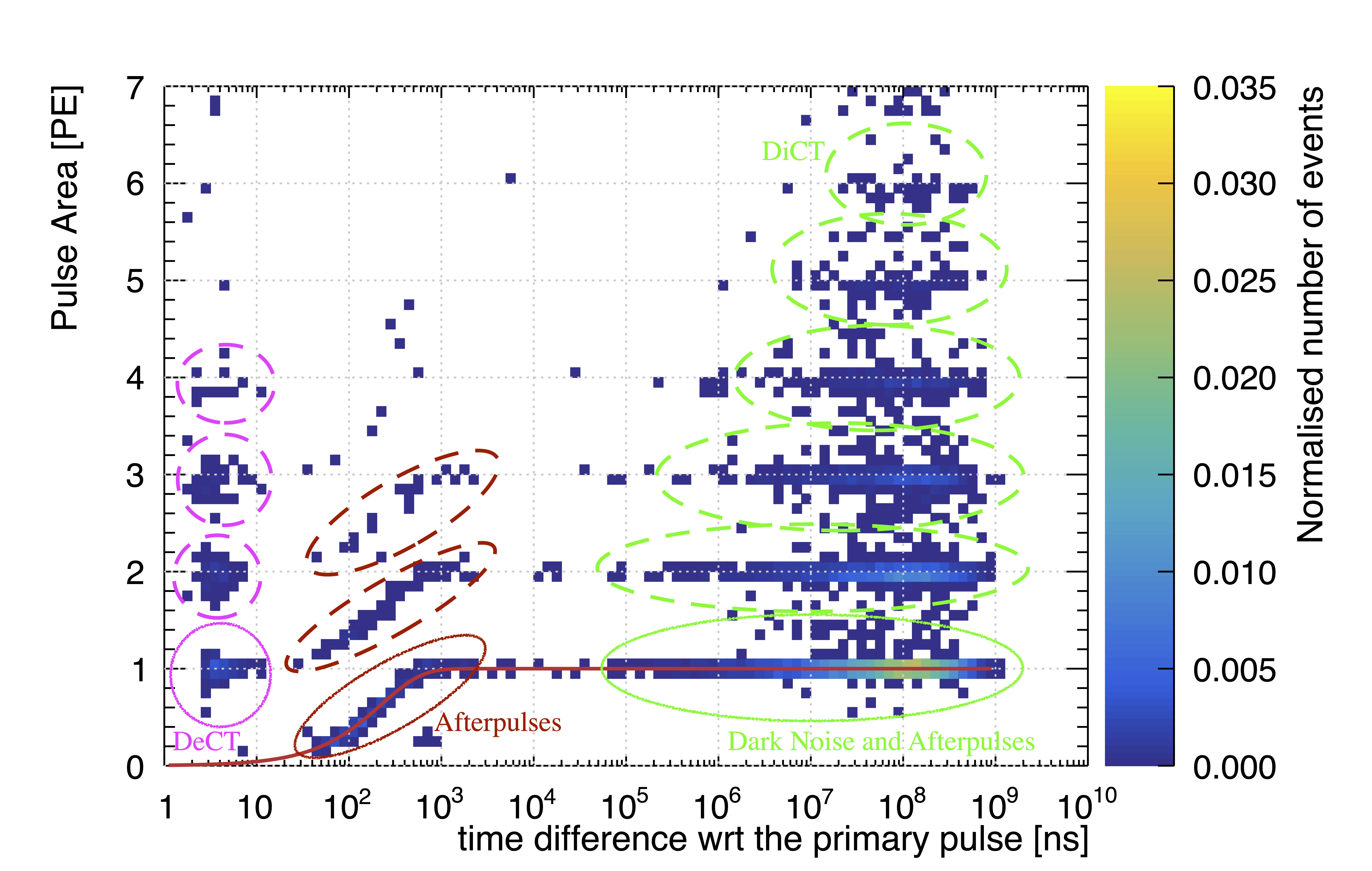}
\includegraphics[width=0.99\linewidth]{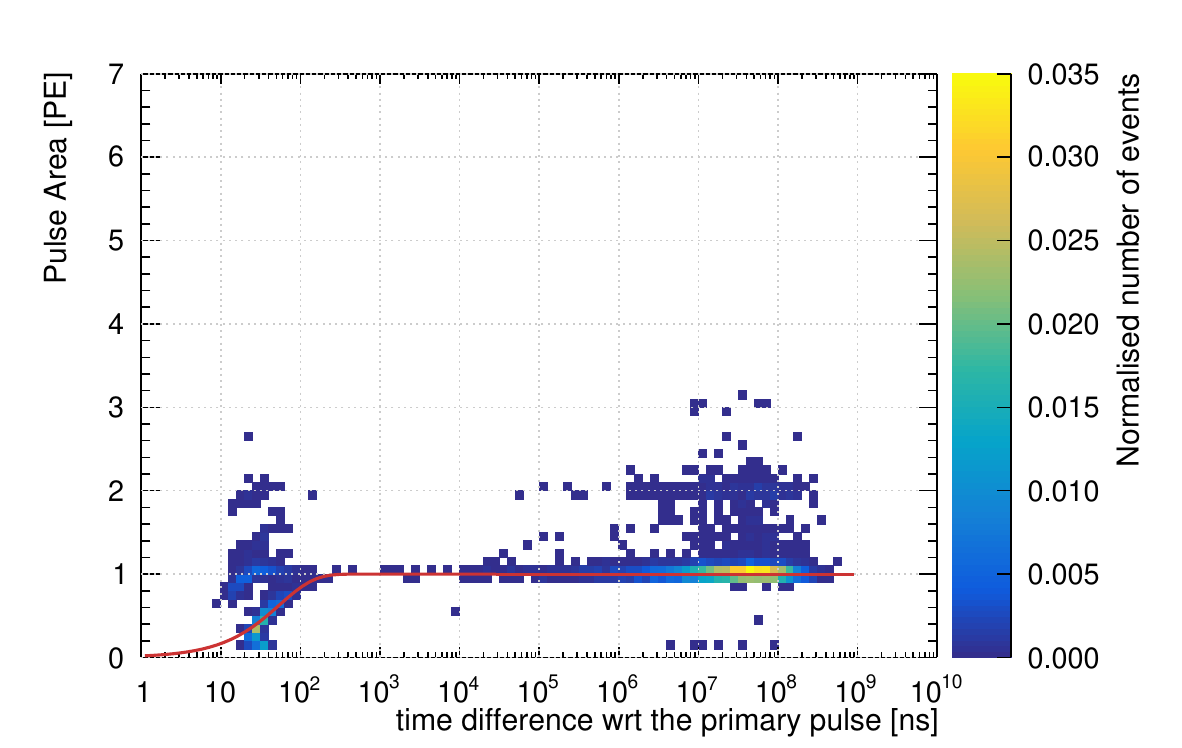}
\caption{FBK~VUVHD3 (top) and HPK VUV4-Q-50 (bottom) charge distribution of first pulses following single PE dark noise driven trigger pulses as a function of the time difference with respect to (wrt) their primary pulse for a temperature of 163~K and for an over voltage of $5.64\pm0.17$~V and $5.37\pm0.16$~V, respectively. The coloured scale in both figure represents the normalised number of events in each bin. In the top figure are also reported the distribution of  photoelectrons for  subsequent  pulses.  See  text  for  detailed  explanation.}
\label{fig:2DHD3VUV4}
\end{figure}
\\For reference, in Fig.~\ref{fig:2DHD3VUV4} we show the charge distribution of first pulses following single PE primary pulses (also referred to as trigger or prompt pulses) obtained with the pulse fitting scheme presented in Sec.~\ref{S:fitsection} for two of the photosensors tested in this work, as a function of the time difference from their primary pulse. These measurements, performed at TRIUMF, were recorded for a temperature of 163~K at an over voltage of $5.64\pm0.17$~V and $5.37\pm0.16$~V for the FBK VUVHD3 SiPM and HPK VUV4-Q-50 MPPC, respectively.\\This representation  allows  the  identification  of  the SiPM dark and correlated signals, just discussed. The  microcells’  recharging  process  and  effects  contributing to  CAs  including  DiCT, DeCT and  afterpulsing  can also be identified. Pulses occurring with a large delay from the parent pulse can originate from either afterpulsing or dark noise (green solid circle) and can suffer DiCT (green dashed circle). Pulses  occurring  shortly  after a  prompt pulse  in  a previously triggered microcell, which are typically afterpulses, have lower charges since the microcell charge has not been fully recharged (red solid circle). They can produce  DiCT events (red dashed circle). Avalanches due to DeCT (pink solid circle) are instead  identified  as  additional  pulses  with  1 PE charge. They can produce DiCT events (pink dashed circle). The solid red line on both figures shows a fit of the afterpulsing events with $\overline{\text{Q}}_{1\text{ PE}}\times\big(1-e^{-\frac{t}{\tau_S}}\big)$ and is used to measure the recovery time $\tau_S$ of one SiPM cell measured to be $225\pm10$ ns for the FBK VUVHD3 SiPM and $55\pm5$ ns for the HPK VUV4-Q-50 MPPC. $\overline{\text{Q}}_{1\text{ PE}}$ is the average single PE charge defined in Sec.~\ref{S:gain}. The difference in the recovery time constant $\tau_S$ of the two photosensors is mainly due their quenching material: polysilicon for FBK and metal for HPK as shown in \cite{Falcone2021} and \cite{Hamamatsu_doc}, respectively.\\\indent Regardless of the mechanism responsible for the CAs generation, all CAs add an extra charge (measurable in PE-equivalent units) in the nEXO acquisition window, expected to extend up to 1$\upmu \text{s}$ after the trigger pulse. This extra charge artificially increases the total number of apparent photons detected by the SiPMs and, more importantly, provides additional event-by-event fluctuations in the total collected charge. In order to reach the nEXO design energy resolution, the SiPM Correlated Avalanche Fluctuation (CAF), defined as the ratio  between the root mean square error  $\sigma_{\Lambda}$  and the average extra charge $\langle\Lambda\rangle$ produced by CAs per pulse, needs to be within a time window of $1\upmu$s after the primary pulse smaller than $0.4$. The parameter relevant to reaching the required nEXO energy resolution is therefore defined as follows:
\begin{linenomath}
\begin{equation}
\label{eq:ratio_CA}
\text{CAF}\equiv\frac{\sigma_{\Lambda}}{1+\langle\Lambda\rangle}
\end{equation}
\end{linenomath}
This specification only refers to CAs produced within the same SiPM and essentially set a maximum limit on the fluctuation of the CAs per pulse on an avalanche-to-avalanche basis. This requirement is directly connected to the model of the energy resolution presented in Sec.~\ref{S:energyresolution}. The CAF is discussed in Sec.~\ref{S:CA}, for all the devices tested. The external crosstalk contribution to the predicted nEXO energy resolution is separately discussed in Sec.~\ref{S:energyresolution}.

\subsubsection{Correlated Avalanche Fluctuation}
\label{S:CA}
\begin{figure}[ht]
\centering\includegraphics[width=0.99\linewidth]{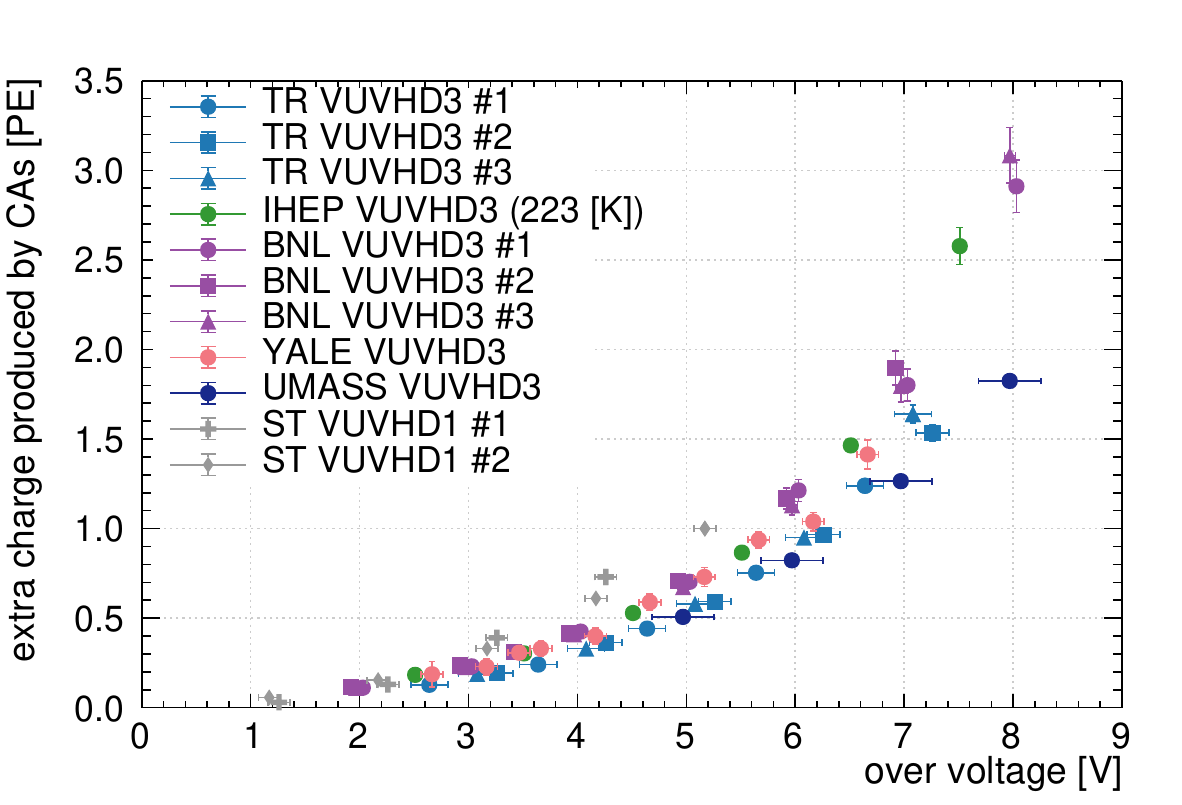}
\caption{Average extra charge produced by Correlated Avalanches (CAs,$\langle\Lambda\rangle$) per primary pulse within a time window of $1\upmu$s after the trigger pulse 
measured at 163~K and as a function of the applied over voltage for FBK VUVHD3 SiPMs. ST VUVHD1 is instead the average extra charge produced by CAs for the previous generation of FBK devices (FBK VUVHD1)~\cite{Ako}.}
\label{fig:CA_mean_HD3}
\end{figure}

\begin{figure}[ht]
\centering\includegraphics[width=0.99\linewidth]{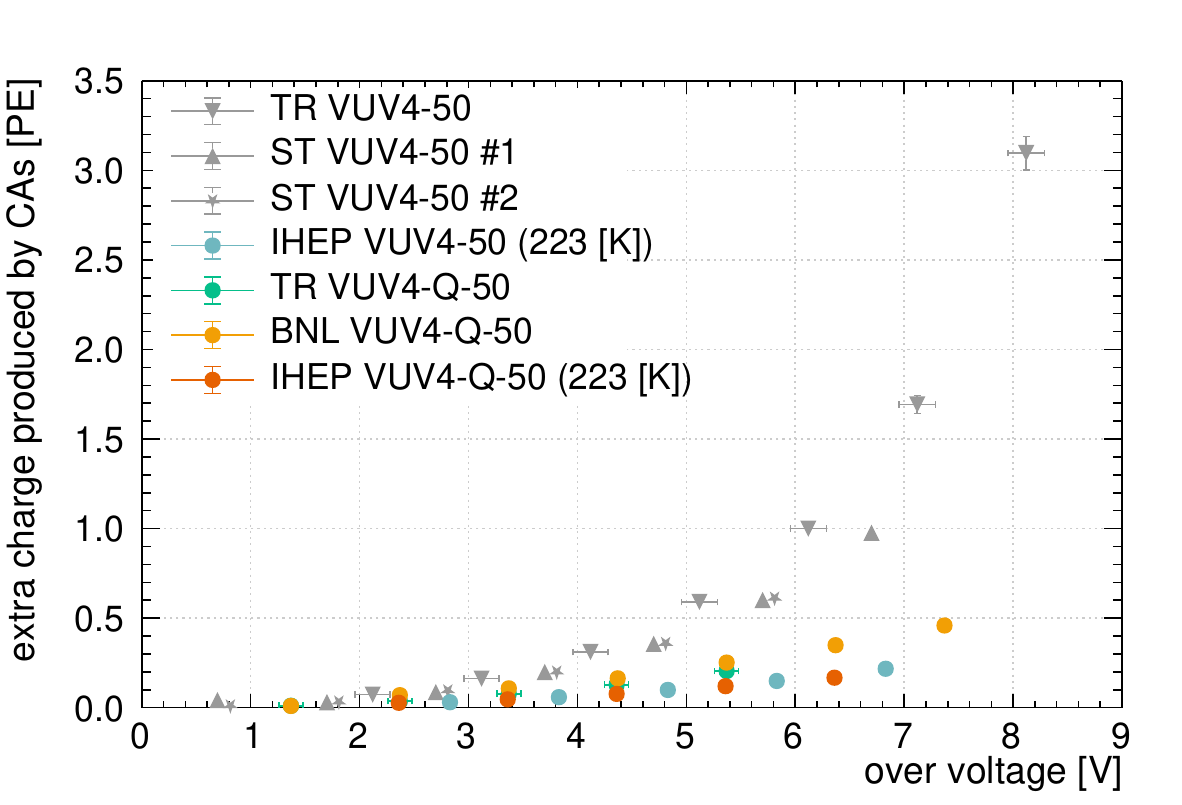}
\caption{Average extra charge produced by Correlated Avalanches (CAs,$\langle\Lambda\rangle$) per primary pulse within a time window of $1\upmu$s after the trigger pulse measured at 163~K and as a function of the applied over voltage for HPK VUV4 MPPCs. TR VUV4-50 and ST VUV4-50 are instead the average extra charge produced by CAs of the previously characterised HPK VUV4-50 MPPCs~\cite{Gallina2019}.}
\label{fig:CA_mean_VUV4}
\end{figure}

\begin{figure}[ht]
\centering\includegraphics[width=0.99\linewidth]{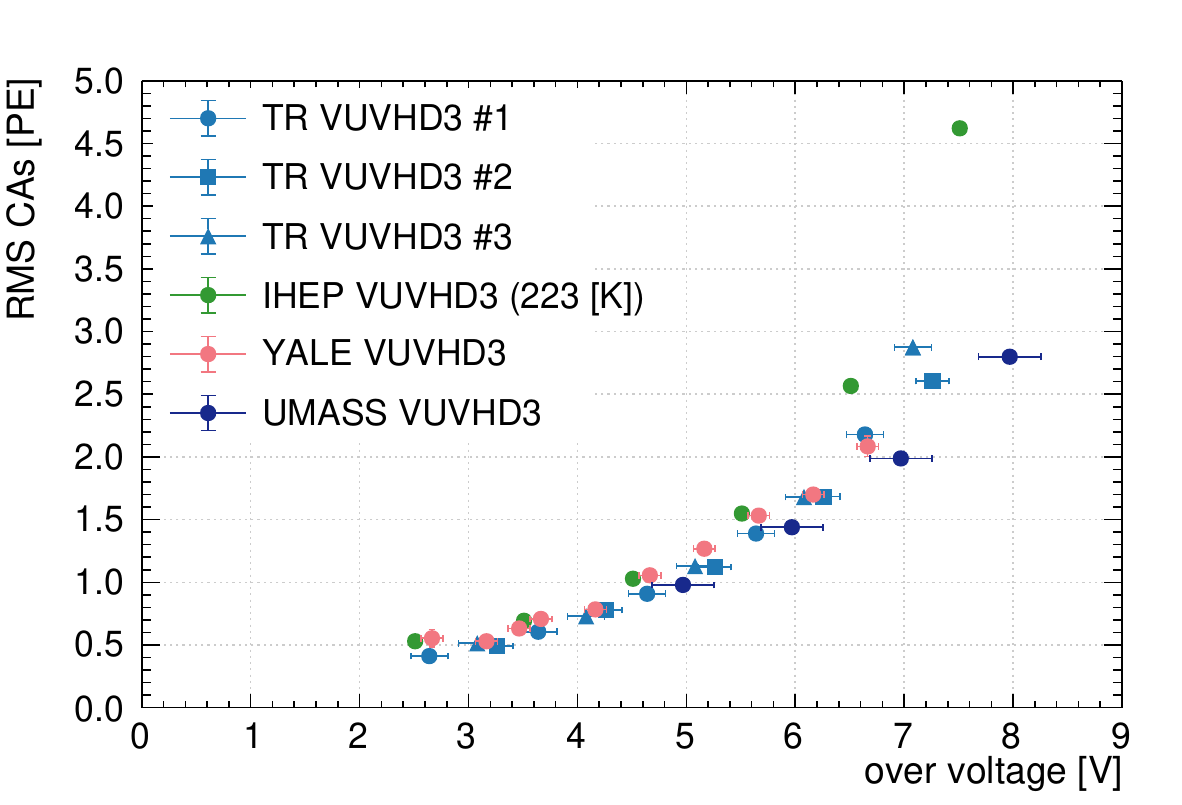}
\caption{Root Mean Square Error (RMS, $\sigma_{\Lambda}$) of the extra charge produced by Correlated Avalanches (CAs) per primary pulse measured at 163~K within a time window of $1\upmu$s after the trigger pulse as a function of the applied over voltage for FBK VUVHD3 SiPMs.}
\label{fig:CA_RMS_HD3}
\end{figure}

\begin{figure}[ht]
\centering\includegraphics[width=0.99\linewidth]{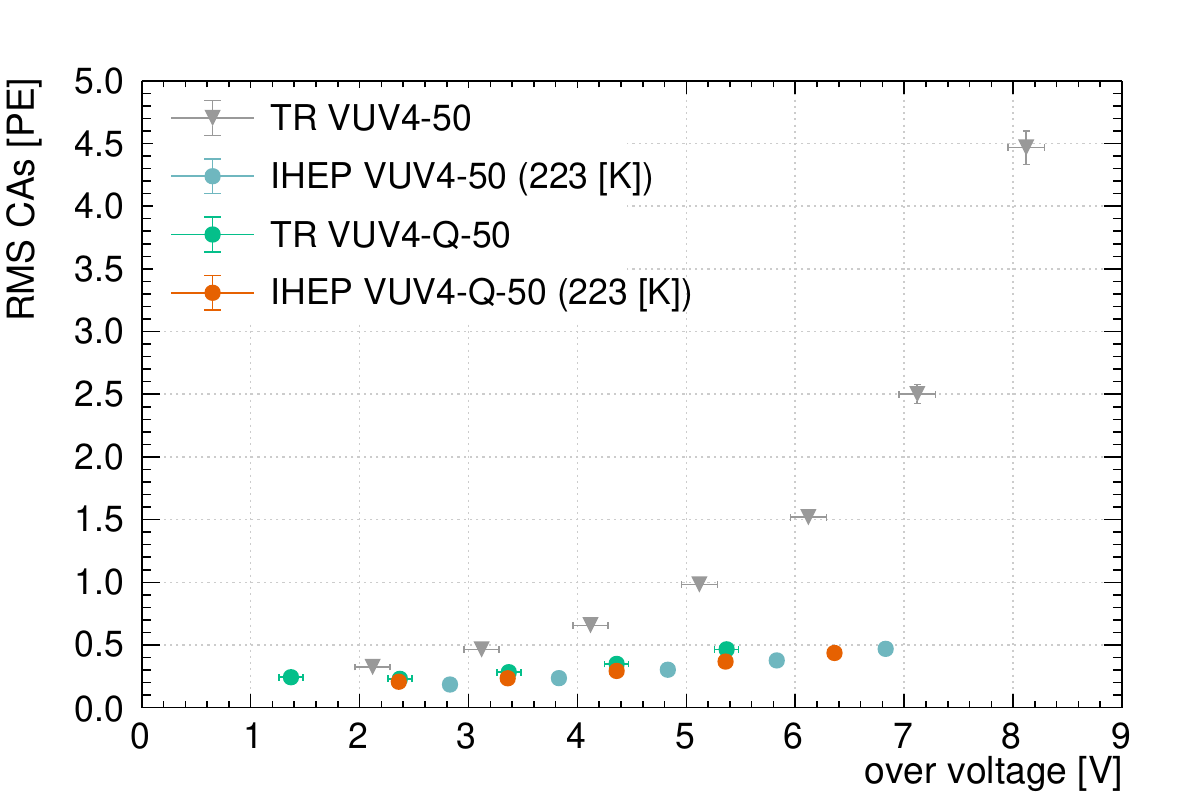}
\caption{Root Mean Square Error (RMS, $\sigma_{\Lambda}$) of the extra charge produced by Correlated Avalanches (CAs) per primary pulse measured at 163~K within a time window of $1\upmu$s after the trigger pulse as a function of the applied over voltage for HPK VUV4s MPPCs. TR VUV4-50 is instead the RMS of the previously characterised HPK VUV4-50 MPPC~\cite{Gallina2019}.}
\label{fig:CA_RMS_VUV4}
\end{figure}

\begin{figure}[ht]
\centering\includegraphics[width=0.99\linewidth]{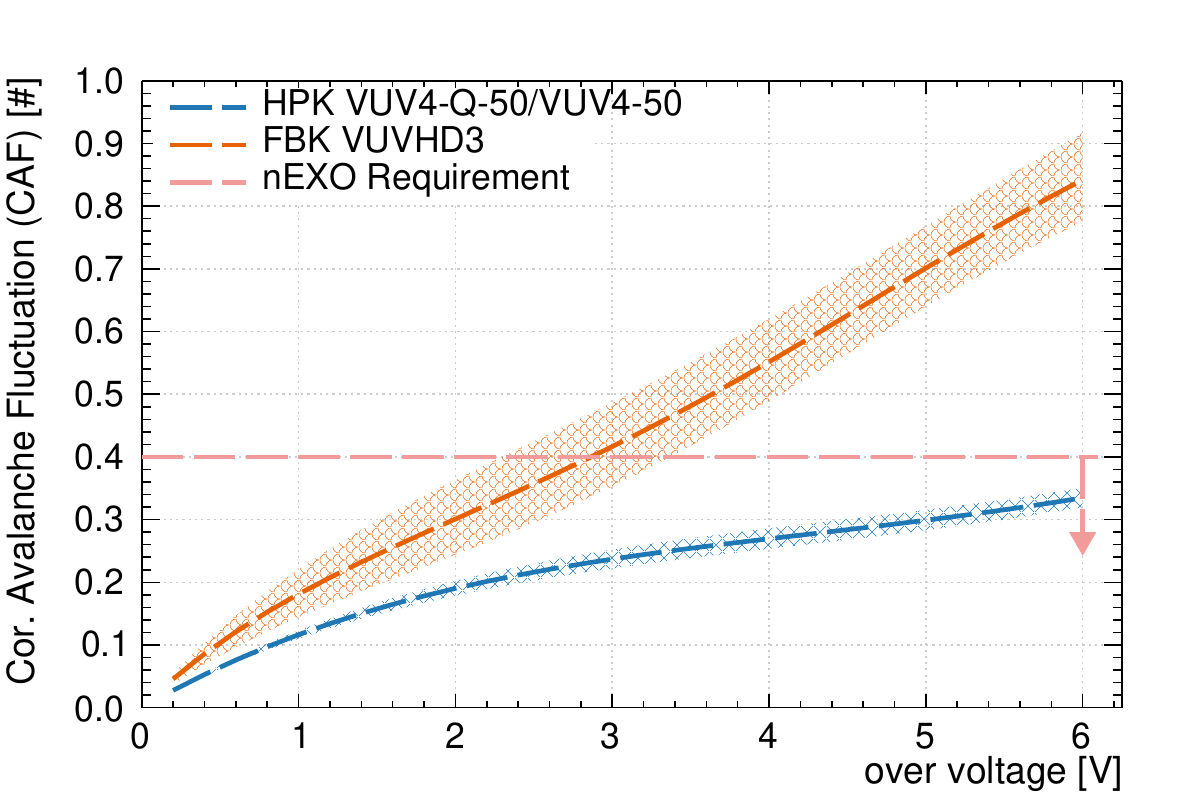}
\caption{Correlated Avalanche Fluctuation (CAF) measured in the  $1\upmu\text{s}$ window after the trigger pulse  as a function of the applied over voltage and at 163~K, as defined by Eq.~\ref{eq:ratio_CA}. The shaded regions represent the spread between all the measurements and were computed interpolating the upper and lower boundaries of the measured data with the corresponding errors. The dashed horizontal line represents the nEXO requirement.}
\label{fig:ratio}
\end{figure}

To evaluate Eq.~\ref{eq:ratio_CA} it is necessary to separately measure $\sigma_{\Lambda}$ and $\langle\Lambda\rangle$ for all the tested photosensors.\\\indent The average extra charge produced by CA per primary pulse, $\langle\Lambda\rangle$, is measured by constructing a histogram of the baseline subtracted waveforms integrated up to $1\upmu\text{s}$ after the trigger pulse and normalized to the average charge of 1 PE pulses, as discussed in Ref.~\cite{Gallina2019}. Waveforms were collected in the dark, with the SiPM shielded from any light source. The average extra charge produced by CAs per pulse is reported in units of PE as a function of the over voltage at 163~K in Fig.~\ref{fig:CA_mean_HD3} for FBK VUVHD3 SiPMs and in Fig.~\ref{fig:CA_mean_VUV4} for HPK VUV4 MPPCs. For comparison, in the same figures we also show previous values of $\langle\Lambda\rangle$ measured at 163~K for the earlier generation FBK VUVHD1 devices~\cite{Ako} and for HPK VUV4 MPPCs reported in Ref.~\cite{Gallina2019}.
IHEP measurements were done at 223~K due to a limitation of their measurement setup. The temperature dependence of the SiPM CA is however expected to be weak, especially within 60~K, as shown in Ref.~\cite{Gallina2019}.\\The new devices show an overall reduction in the average extra charge produced by CAs per pulse. This improvement is particularly significant for HPK MPPCs. For instance, at 3~V of over voltage and 163~K, we measure 
$\langle\Lambda\rangle = 0.23\pm0.06$~PE for FBK VUVHD3 SiPMs
and $\langle\Lambda\rangle = 0.06\pm0.02$~PE for HPK VUV4 MPPCs\footnote{3~V of over voltage is the highest over voltage point for which: (i) the energy resolution of HPK MPPCs and FBK SiPMs is close to its minimum (Sec.~\ref{S:energyresolution}), (ii) all the nEXO requirements (Sec.~\ref{S:intro}) are satisfied, within errors.}\footnote{These values were obtained by using polynomial spline interpolations (forced to go to zero at 0~V of over voltage) of all the data of Fig.~\ref{fig:CA_mean_HD3} for FBK SiPMs and Fig.~\ref{fig:CA_mean_VUV4} for HPK VUV4 MPPCs.  Similar spline interpolations were also used to compute the reference values of other measured quantities ({\it e.g.} RMS ($\sigma_{\Lambda}$), dark count rate, PDE etc ..), always at 3~V of over voltage. Moreover in consideration of the significant improvement of HPK MPPCs as compared to the ones previously tested in Ref.~\cite{Gallina2019}, we used only the new measurements to compute $\langle\Lambda\rangle$, $\sigma_{\Lambda}$ and the corresponding CAF for HPK MPPCs.}.\\
\indent Fig.~\ref{fig:CA_RMS_HD3} and Fig.~\ref{fig:CA_RMS_VUV4} show the Root Mean Square (RMS) error, $\sigma_{\Lambda}$, of the extra charge produced by CAs per primary pulse, as a function of the applied over voltage, measured at 163~K and in units of PE for FBK VUVHD3 SiPMs and HPK VUV4 MPPCs, respectively. IHEP measurements were performed at 223~K. The RMS is however expected to have a negligible temperature dependence, as shown in Ref.~\cite{gallina2021development}.
In general, the RMS has a sharper increase with over voltage compared to the mean. This results in a significant fluctuation of the extra charge produced by CAs on an avalanche by avalanche basis. Moreover, the HPK MPPCs tested for this work have a smaller RMS with respect to both FBK VUVHD3 SiPMs and the previously tested HPK MPPCs \cite{Gallina2019}. This behavior is in agreement with the corresponding trend of $\langle\Lambda\rangle$ for these devices (Fig.~\ref{fig:CA_mean_VUV4}). At 163~K and 3~V of over voltage the RMS is $0.51\pm0.06$~PE and $0.25\pm0.01$~PE for FBK VUVHD3 SiPMs and HPK VUV4 MPCCs, respectively.\\\indent Fig.~\ref{fig:ratio} shows the CAF obtained with Eq.~\ref{eq:ratio_CA} by using a polynomial spline interpolation (forced to go to zero at 0~V of over voltage) of all the data of Figs.~\ref{fig:CA_mean_HD3},~\ref{fig:CA_RMS_HD3} for FBK SiPMs and Figs.~\ref{fig:CA_mean_VUV4},~\ref{fig:CA_RMS_VUV4} for HPK MPPCs. The nEXO requirement is also shown in the same figure. The shaded regions represent the spread between all the measurements and were computed interpolating the upper and lower boundaries of the measured data with the corresponding errors. Given that these devices are physically different pieces, not pre-selected before testing, they represent a conservative estimate of the uncertainty of the final nEXO production.\\For reference, at 3~V over voltage and a temperature of 163~K, the CAF is equal to  $0.42\pm0.07$ for FBK VUVHD3 SiPMs and to $0.24\pm0.02$ for HPK VUV4 MPPCs. Overall the smaller extra charge produced by CAs per pulse allows for an increase in the photosensors operational over voltage, as compared with the ones previously tested. This allows higher single PE gain (Sec.~\ref{S:gain}) and, in turn, better signal to noise ratio of the nEXO photon detection system.

\subsubsection{Dark Count Rate and Number of Correlated Delayed avalanches}
\label{S:DN}

\begin{figure}[ht]
\centering\includegraphics[width=0.99\linewidth]{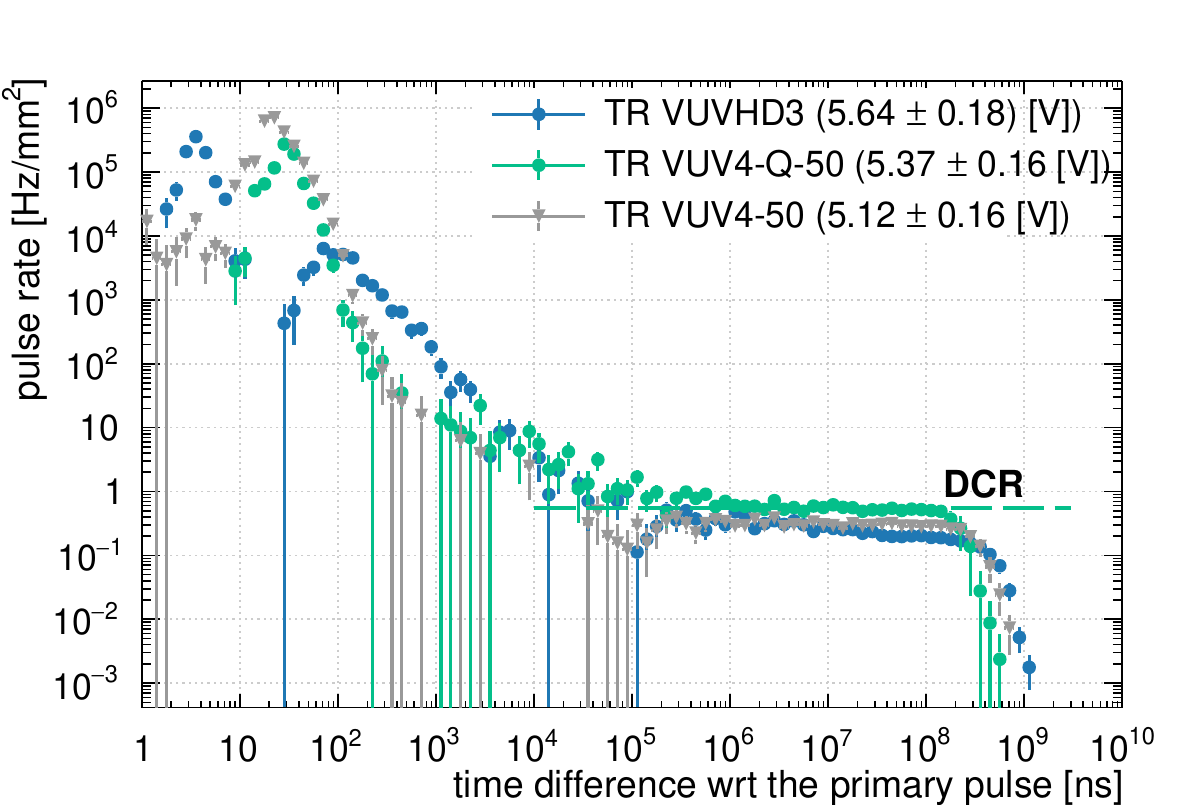}
\caption{Observed dark pulse rate $\text{R}(t)$ measured for the FBK~VUVHD3 SiPM and for HPK VUV4-Q-50 MPPC as a function of the time difference with respect to (wrt) the primary pulse at $163~\text{K}$ and for roughly 5~V of over voltage. The horizontal line represents the DCR extrapolated from the pulse rate of the HPK VUV4-Q-50 MPPC. TR VUV4-50 represents the observed dark pulse rate $\text{R}(t)$, always measured at 163~K and as a function of the time difference with respect to the primary pulse for the previously characterised HPK VUV4-50 MPPC~\cite{Gallina2019}.
}
\label{fig:pulserate}
\end{figure}

Dark noise pulses can be distinguished from CDA pulses by studying their time distribution relative to the primary pulse, as shown in Ref.~\cite{Butcher2017}. The secondary pulse rate, $\text{R}(t)$, is computed as a function of the time difference, $t$, from the primary pulse ($t=0$) as:
\begin{linenomath}
\begin{equation}\label{eq:rt}
    \text{R}(t) = \text{R}_{\text{DCR}}(t)+ \text{R}_{\text{CDA}}(t)
\end{equation}
\end{linenomath}
where $\text{R}_{\text{DCR}}$ is the Dark Count Rate (DCR) and $\text{R}_{\text{CDA}}$ is the rate of the CDAs per pulse. The measured $\text{R}(t)$ at 163~K and for roughly 5~V over voltage is reported for FBK and HPK devices in Fig.~\ref{fig:pulserate}. 

\begin{figure}[ht]
\centering\includegraphics[width=0.99\linewidth]{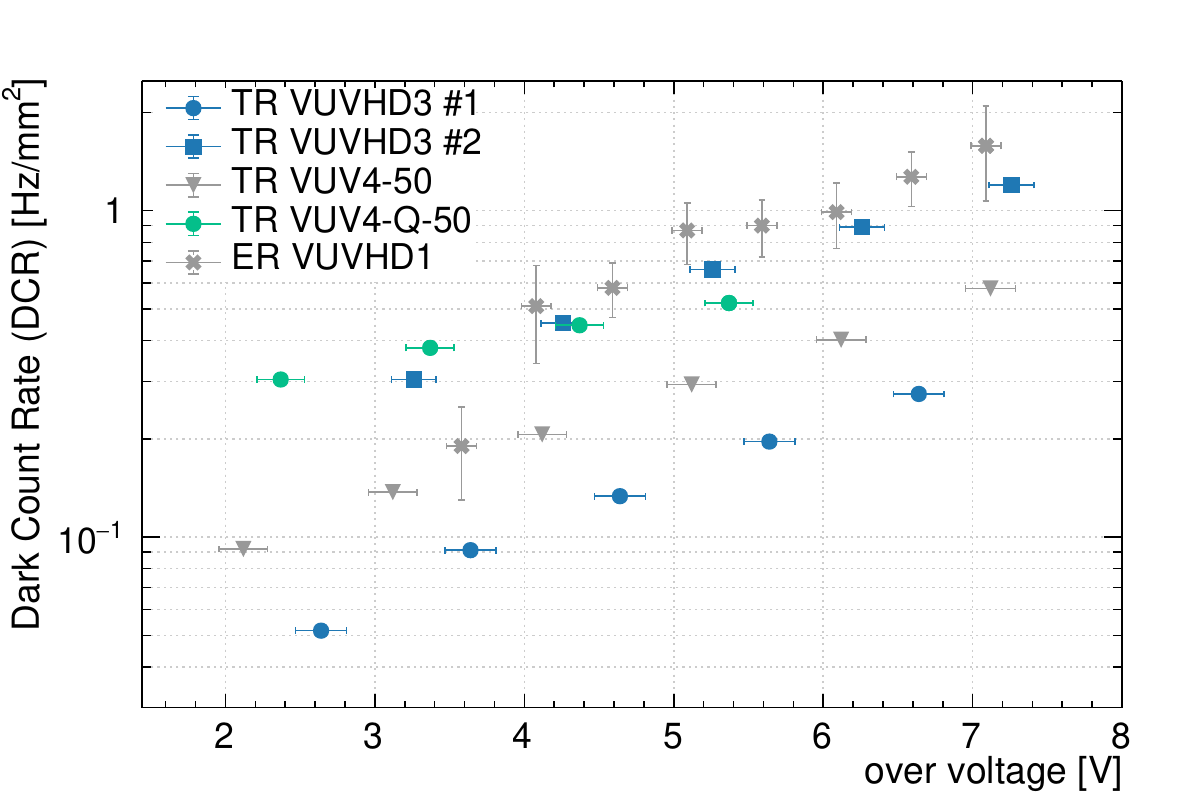}
\caption{Dark Count Rate (DCR) normalized by the SiPM photon sensitive area, as a function of the applied over voltage, measured at 163~K. ER VUVHD1 and TR VUV4-50 are 
the DCR, always  normalized by the SiPM photon sensitive area, for the previously characterised FBK VUVHD1 and  HPK VUV4-50 MPPC~\cite{Ako,Gallina2019}. The nEXO requirement is out of the range of plot.}
\label{F:DN_OV}
\end{figure}

\begin{figure}[ht]
\centering\includegraphics[width=0.99\linewidth]{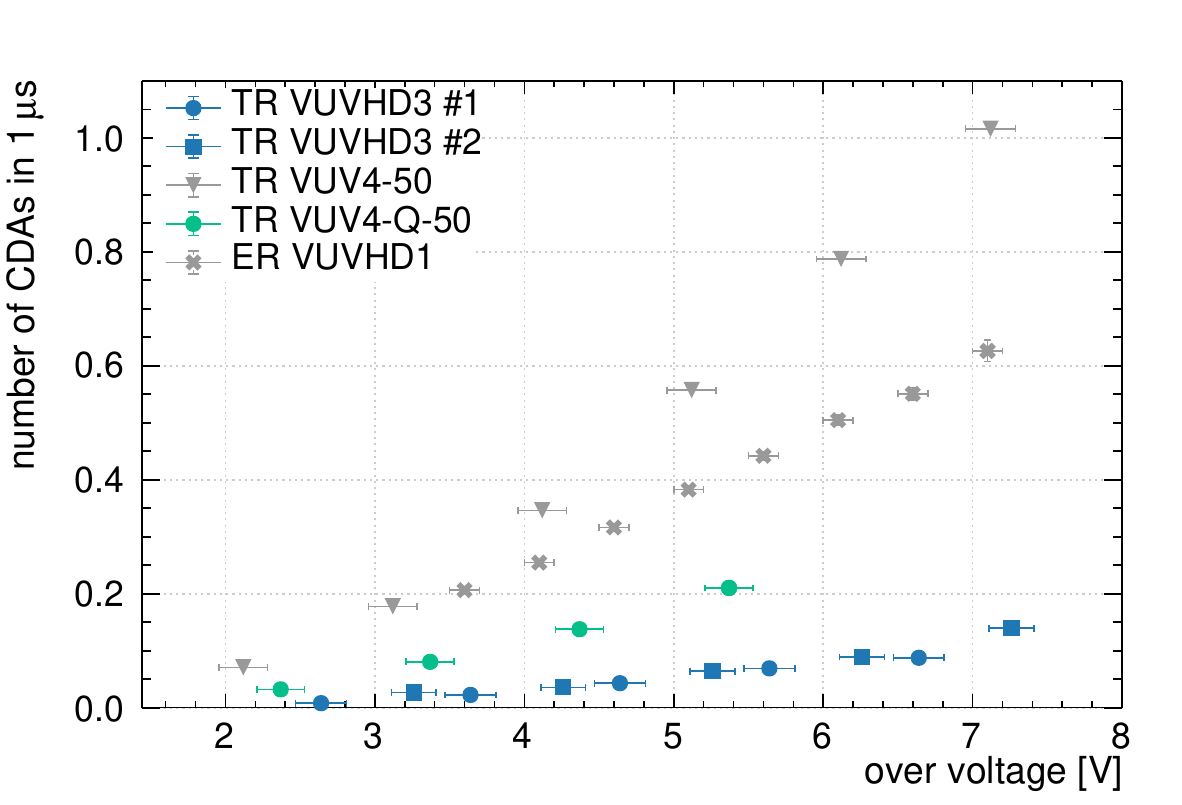}
\caption{Number of Correlated Delayed Avalanches (CDAs) per primary pulse within a time window of 1 $\upmu$s after the trigger pulse, as a function of the applied over voltage, measured at 163~K. ER VUVHD1 and TR VUV4-50 are the number of CDAs  for the previously characterised FBK VUVHD1 and  HPK VUV4-50 MPPC~\cite{Ako,Gallina2019}.}
\label{F:AP_OV}
\end{figure}

The shape of the pulse rate Time Distribution (TD) depends generally on several factors. At a short time ($<10^3$~ns) the TD is dominated by afterpulses and DeCT. All the curves show the presence of a maximum due to afterpulse because afterpulses are either not present at short times or they start to be resolved from their primary pulse only when the time $t$ at which they happen is comparable with the SiPM SPAD recovery time constant. The TD of the FBK VUVHD3 SiPM shows two maxima. The one at later times ($\sim10^2$~ns) is due to afterpulse, explained above. The one instead at $\sim3$~ns is due to DeCT. This can be seen combining Fig.~\ref{fig:pulserate} with, for example, Fig.~\ref{fig:2DHD3VUV4} that shows a significant DeCT component at short times for the FBK VUVHD3 SiPM after the primary pulse. At a long time ($>10^6$~ns)  the TD shows instead its uncorrelated component (DCR) and it flattens because the DCR is not correlated with the primary pulse. The SiPM DCR at 163~K 
can therefore be obtained, as a function of the applied over voltage, from Fig.~\ref{fig:pulserate} by performing a weighted mean of the pulse rates in the range [$10^7,10^8$]~ns.  DCR errors are assessed as standard errors.\\\indent The DCR measured at 163~K as a function of the over voltage, for all the SiPMs and MPPCs tested, are shown in Fig.~\ref{F:DN_OV}. In the same figure we also report the DCR for the previous generation of FBK devices (FBK VUVHD1) and HPK VUV4-50 MPPC measured in Ref.~\cite{Ako} and  Ref.~\cite{Gallina2019}, respectively.\\\indent All the SiPMs and MPPCs comfortably satisfy the nEXO requirements ($\le10~\text{Hz}/\text{mm}^2$, Sec.~\ref{S:intro}) with similar performances. We however acknowledge quite a large spread in the photon-sensors DCR, even for SiPMs being part of the same wafer (FBK).\\For instance, at 163~K and 3~V of over voltage we measure an average DCR of $0.35\pm 0.01~\text{Hz/mm}^{2}$  for the HPK VUV4-Q-50 MPPC, close to that of the previously characterised HPK VUV4-50 MPPC \cite{Gallina2019}, and of $0.19\pm0.01~\text{Hz/mm}^{2}$ for FBK VUVHD3 SiPMs.
Fig.~\ref{fig:pulserate} can also be used to estimate the number of correlated delayed avalanches per pulse within a fixed time window of length $\Delta t$. The number of CDAs per pulse can in fact be computed by applying Eq.~\ref{eq:rt} to the observed pulse rate as follows:
\begin{equation}
\label{eq:CDA_eq}
\text{N}_{\text{CDA}}(\Delta t) = \int_{0}^{\Delta t} \Big( \text{R}(t)- \text{R}_{\text{DCR}}(t) \Big)~\text{dt} 
\end{equation}
where $\text{R}_{\text{DCR}}$ is the DCR reported in Fig.~\ref{F:DN_OV}. The measured average number of CDAs per pulse in the $1\upmu$s window after the trigger pulse is reported in Fig.~\ref{F:AP_OV}, as a function of the applied over voltage. HPK MPPCs present, on average, a larger amount of correlated delayed avalanches as compared to FBK~VUVHD3 SiPMs, especially at short times after the trigger pulse ($\le100~\text{ns}$, Fig.~\ref{fig:pulserate}) where the time distribution is dominated by afterpulse events. The high HPK~VUV4 MPPC afterpulse rate has, in fact, already been reported in Ref.~\cite{Gallina2019} where we noted shoulder-like events in the HPK MPPC charge distribution. This has been attributed to fast CAs which do not get resolved from their parent primary pulse by the DAQ~($\le4~\text{ns}$). However, the HPK MPPCs tested in this work featured an almost two-fold reduction in the number of CDAs compared to the MPPC (HPK VUV4-50) tested in Ref.~\cite{Gallina2019}. For $163~\text{K}$ and 3~V of over voltage the average number of CDAs per pulse in $1\upmu$s is found to be $0.060\pm0.003$.

We also record a significant reduction in the number of CDAs for the new generation of FBK devices (FBK~VUVHD3) that results in the lowest measured value for all the devices tested. This is attributed to the triple doping technology developed by FBK, briefly discussed in Sec.~\ref{S:intro} \cite{Capasso2020}. For our reference over voltage and temperature ($163~\text{K}$, 3~V) the average number of CDAs per pulse in $1\upmu$s is $0.017\pm0.001$. Finally, it is worth noting that the number of CDAs extrapolated in this section cannot be compared directly with the average extra charge per pulse produced by CAs reported in Sec.~\ref{S:CA}. The number of CDAs is in fact derived according to Ref.~\cite{Butcher2017}, which takes into account only the time differences of delayed events with respect to their primary pulse. The average extra charge produced by CAs, on the other hand, takes into consideration their different charges as well.

\subsubsection{Number of Additional Prompt Avalanches}
\label{S:CT}

\begin{figure}[ht]
\centering
\includegraphics[width=0.99\linewidth]{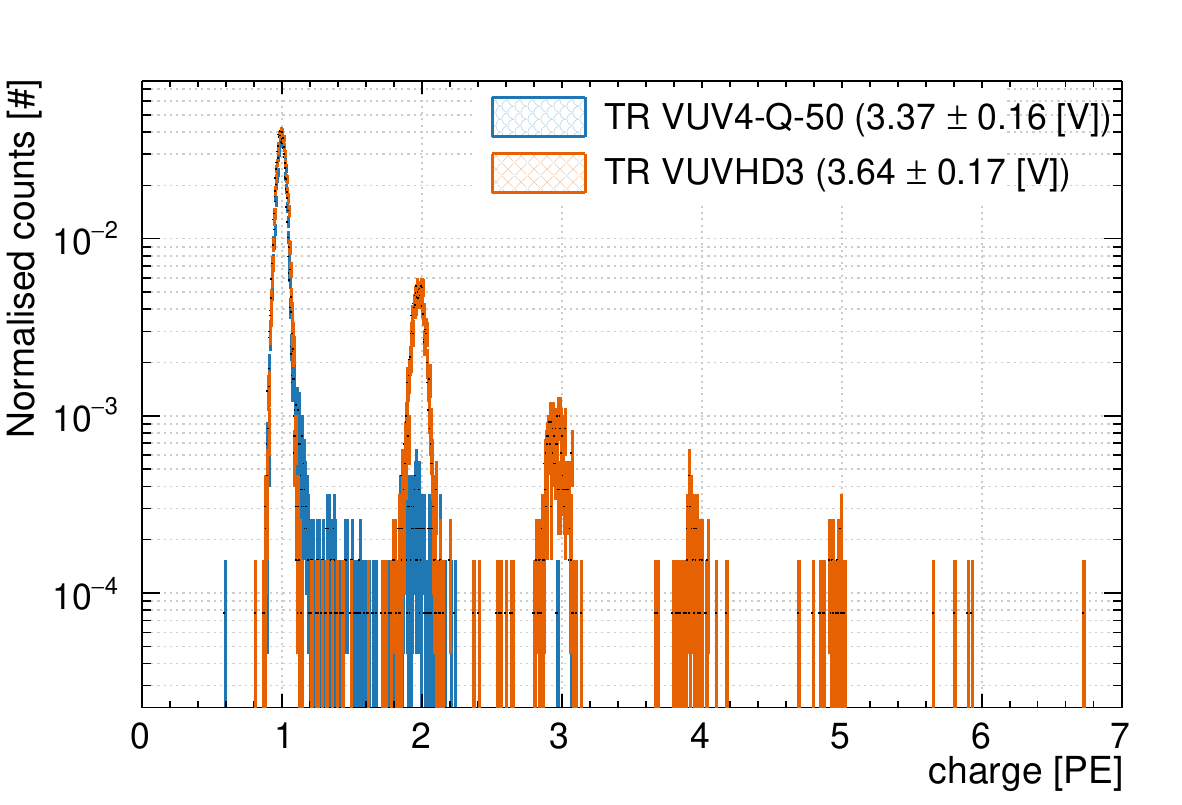}
\caption{Charge distribution for prompt pulses obtained using dark data for FBK VUVHD3 SiPMs and HPK VUV4-Q-50 MPPC measured for roughly 3.5~V of over voltage and for a temperature of 163~K.}
\label{fig:charge}
\end{figure}

\begin{figure}[ht]
\centering\includegraphics[width=0.99\linewidth]{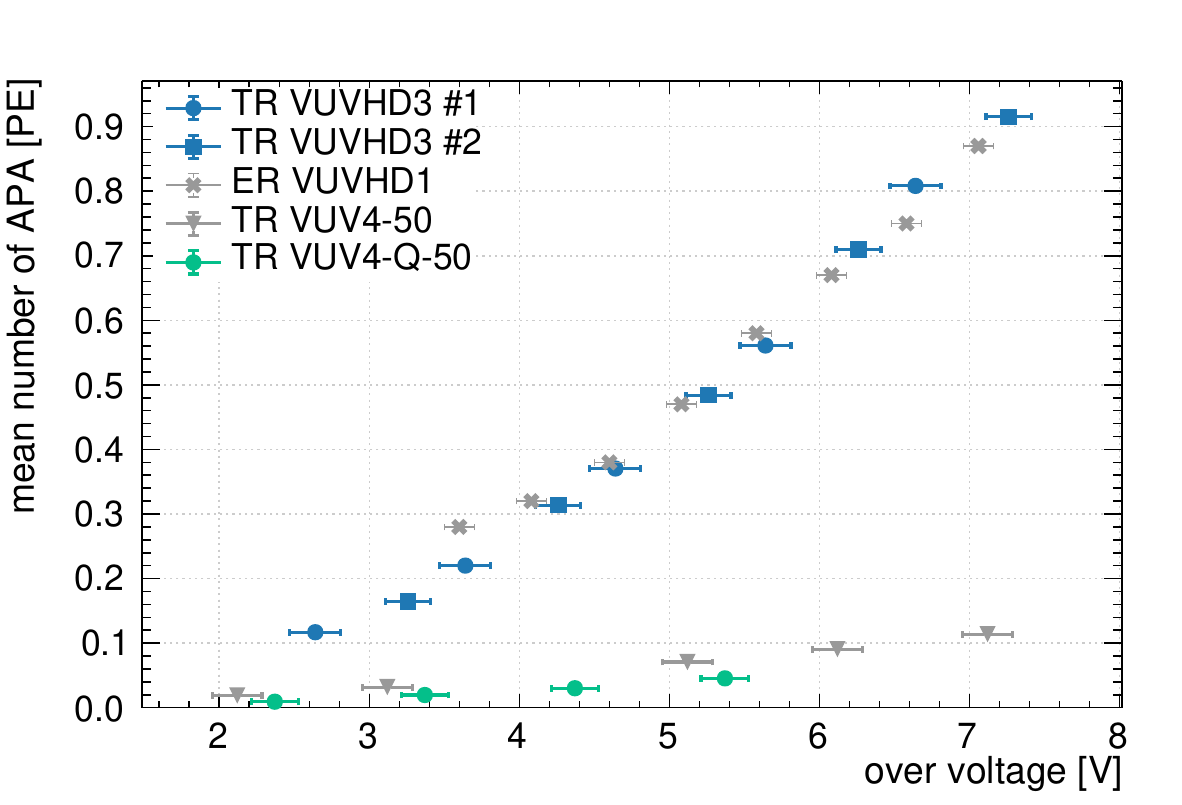}
\caption{Number of Additional Prompt Avalanches (APAs) measured at 163~K as a function of the over voltage for all the devices tested in this work. ER VUVHD1 and TR VUV4-50 are the Number of APAs for the previously characterised FBK VUVHD1 and  HPK VUV4-50 MPPC~\cite{Ako,Gallina2019}.}
\label{fig:APA}
\end{figure}

Based on the measured dark noise rate reported in Sec.~\ref{S:DN}, and assuming Poisson statistics, the probability of having two dark noise pulses occurring within a few nano-seconds is negligible. Therefore, the collected dark data can be used to investigate DiCT, as shown in Ref.~\cite{Gallina2019}.\\DiCT occurs when photons generated during an avalanche in one micro-cell promptly travel to the amplification region of neighbouring micro-cells, where they induce a secondary avalanche. This mechanism mimics a multiple PE signal, thus biasing the photon counting ability of the device.\\The charge distribution of the prompt pulses\footnote{We define a prompt pulse as the first SiPM pulse in each waveform recorded by the DAQ} obtained from the dark data with the photosensors set at different over voltages and 163~K, can then be used to determine the mean number of Additional Prompt Avalanches (APA)s, $\text{N}_{\text{APA}}$, due to Direct CrossTalk as follows:
\begin{linenomath}
\begin{equation}
\text{N}_{\text{APA}}=\frac{1}{N}\sum_{i=1}^{N}\frac{\text{Q}_i}{\overline{\text{Q}}_{\text{1 PE}}}-1
\end{equation}
\end{linenomath}
where $\text{Q}_i$ is the charge of the prompt pulse $i$, $\overline{\text{Q}}_{\text{1 PE}}$ is the average charge of a single PE pulse, defined in Sec.~\ref{S:gain}, and $N$ is the number of prompt avalanches analyzed. An example of prompt pulse charge distribution for roughly 3.5~V of over voltage is shown in Fig.~\ref{fig:charge}. 
The $\text{N}_{\text{APA}}$ number, in units of PE, measured at a temperature of 163~K and as a function of the applied over voltage is reported in Fig.~\ref{fig:APA} for all the devices tested. 
The number of APAs for FBK SiPMs is significantly larger than for HPK MPPCs, most likely due to the different SPADs isolation material: polysilicon for FBK SiPMs compared to metal, probably tungsten, for HPK MPPCs~\cite{Merzi2022}. Overall, neither FBK SiPM nor HPK MPPCs show significant improvement from previously tested ones \cite{Ako,Gallina2019}. From this we can conclude that the reduced extra charge produced by CAs of the HPK MPPCs and FBK SiPMs tested for this work and shown in Sec.~\ref{S:CA}, is mainly due to a reduction in their afterpulses (Sec.~\ref{S:DN}), rather than a suppression of their direct optical cross talk. For $163~\text{K}$ and 3~V of over voltage we measure a mean number of APAs from DiCT of $0.148\pm 0.003$~PE for FBK VUVHD3 SiPMs and $0.016\pm0.002$~PE for HPK VUV4-Q-50 MPPCs. The small amount of $\text{N}_{\text{APA}}$ can also be understood looking at Fig.~\ref{fig:charge}  where 3 PE events are greatly suppressed for HPK MPPCs.

\subsection{Photon Detection Efficiency}
\label{S:PDE}

The PDE is the combined 
probability that a photon is absorbed in the active volume of the SiPM with a subsequently triggered avalanche~\cite{Gallina2019avalanche}. To meet the nEXO requirements, the PDE must be $\geq15\%$ for $~175\text{ nm}$. In two previously reported nEXO studies ~\cite{Ako,Gallina2019} the 
PDE of the HPK VUV4 MPCCs and FBK SiPMs was measured in pulse counting mode using a pulsed Xenon light source or a gaseous Xenon scintillation light source. The light flux in both cases was calibrated with a Hamamatsu PMT (HPK R9875P). For this work, the PDE of all the SiPMs and MPPCs was measured using continuous lamps, vacuum monochromators\footnote{The vacuum monochromators used by TRIUMF and IHEP are: a VM200 Resonance monochromator, and a Vacuum 302 McPherson, respectively.} and NIST calibrated photodiodes (AXUV100G). \\\indent The PDE of the FBK VUVHD3 SiPMs and HPK MPPCs was measured by TRIUMF and IHEP in normal incidence, in the wavelength range [160-200]~nm  at temperatures of 163~K and 300~K, respectively\footnote{The limitations of the IHEP setup prevent the measurement of the photosensor PDE at cryogenic temperatures.}. The temperature dependence of the SiPM PDE is however expected to be weak,  as shown from recently published results \cite{Zhang2022,Acerbi2022}. The procedure involves measurement of the PDE of the devices under test at a fixed wavelength ($175$~nm) and as a function of the applied over voltage. This measurement is carried out using two techniques reported in Sec.~\ref{S:PDE_TRIUMF_tech} and Sec.~\ref{S:PDE_IHEP_tech} that account for the different experimental conditions of the two measurement setups ({\it i.e.} different temperature and, therefore, different SiPM noise characteristics), as shown in Sec.~\ref{S:PDE_174}. The wavelength dependence of the SiPM PDE was then extracted using the measured PDE values at $175$~nm and by constructing a correlation between the SiPM current under illumination and its PDE, as described in Sec.~\ref{S:PDE_wavelength}.

\subsubsection{TRIUMF Technique}
\label{S:PDE_TRIUMF_tech}

\begin{figure}[ht]
\centering\includegraphics[width=0.99\linewidth]{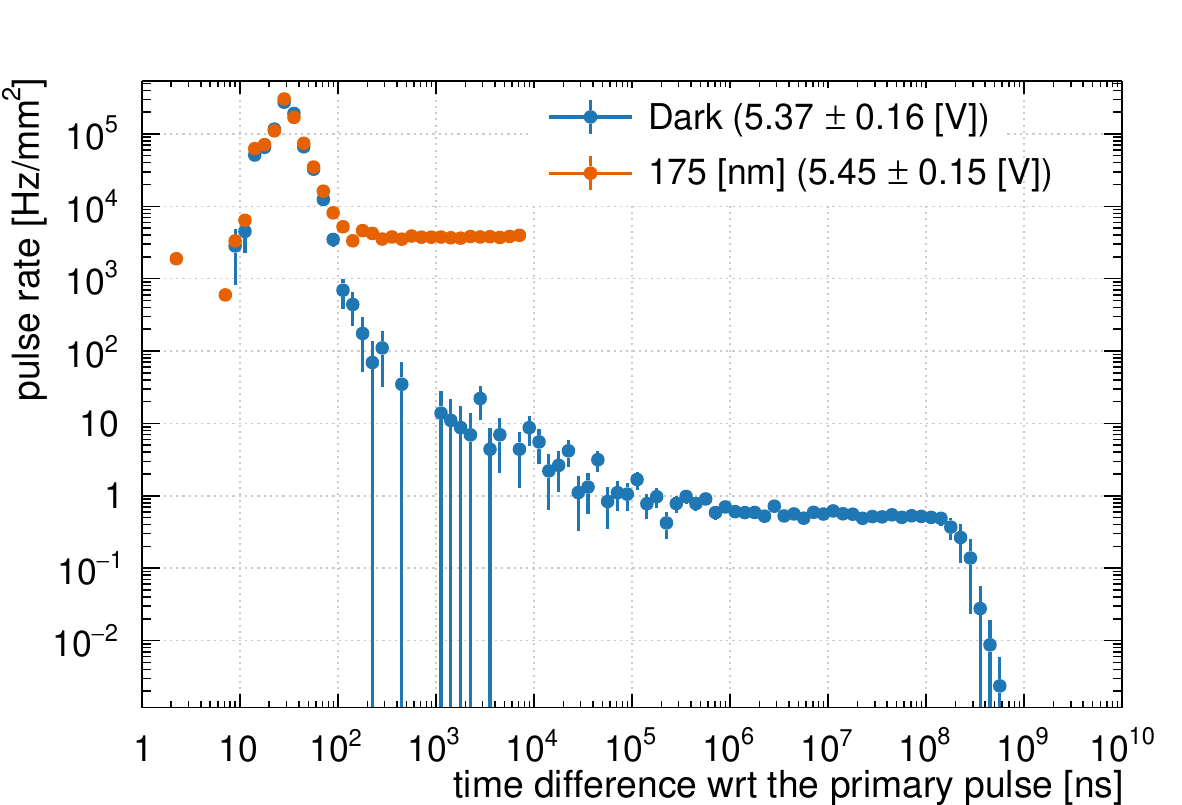}
\caption{Observed pulse rate $\text{R}(t)$ measured at 163~K and for roughly 5~V of over voltage with and without $175$~nm illumination, as a function of the time difference with respect to (wrt) the primary pulse for HPK VUV4-Q-50 MPPC.
}
\label{fig:pulserateMEG2lightdark}
\end{figure}

At TRIUMF, the measurement of the PDE at $175$~nm is based on the estimation of the SiPMs photon induced avalanche rate as a function of the applied over voltage, using the time distribution between pulses, in analogy with what was done in Sec.~\ref{S:DN} to measure the SiPMs DCR. More precisely, since photon induced avalanches are uncorrelated events (like dark noise events), it is possible to distinguish them from dark noise and correlated delayed avalanches by studying the time distribution of all the events relative to the primary pulse. The total observed pulse rate $\text{R}(t)$ can then be computed using Eq.~\ref{eq:rt} with an additional contribution due to photon induced avalanches, as follows:
\begin{linenomath}
\begin{equation}
\label{eq:rt_light}
    \text{R}(t) = \text{R}_{\text{DCR}}(t)+\text{R}_{\text{CDA}}(t)+\text{R}_{\text{0}}(t)
\end{equation}
\end{linenomath}
where: $\text{R}_{\text{0}}(t)$ is the rate of photon induced avalanches detected by the SiPM, $\text{R}_{\text{DCR}}(t)$ the SiPM DCR measured in Sec.~\ref{S:DN}, and $\text{R}_{\text{CDA}}(t)$ the correlated delayed avalanche rate. In Fig.~\ref{fig:pulserateMEG2lightdark}, we report the HPK VUV4-Q-50 pulse rate $\text{R}(t)$ measured at a temperature of 163~K under $175$~nm illumination as a function of the time difference with respect to the primary pulse. 
In the same figure we also show the corresponding pulse rate measured without illumination (dark condition) for roughly the same over voltage, as reported in Fig.~\ref{fig:pulserate}. It is clear that the pulse rate  under illumination is significantly higher due to the photon induced contribution. The uncorrelated pulse rate, ($\text{R}_{\text{DCR}}(t)+ \text{R}_{\text{0}}(t)$) can then be measured, as a function of the over voltage, from Fig.~\ref{fig:pulserateMEG2lightdark}, by performing a weighted mean of the asymptotic pulse rates at long times in analogy with what was done in Sec.~\ref{S:DN} for the DCR. Errors on the measured rates, with and without illumination, are computed as standard errors\footnote{$\text{R}_{\text{0}}(t)$ is extracted by performing a weighted mean of the pulse rates under illumination in the range $[1-5]\times10^3$~ns. $\text{R}_{\text{DCR}}(t)$ is extracted performing a weighted mean of the pulse rates without illumination in the range [$10^7,10^8$] ns, as previously noted in Sec.~\ref{S:DN}.}. 
The asymptotic pulse rate is $\sim$4 orders of magnitude lower without illumination, making the contribution of the DCR at 163~K towards the total pulse rate completely negligible. The SiPM PDE can then be obtained dividing $\text{R}_{\text{0}}(t)$ by the photon flux $\Phi_{0}$, measured with a calibrated diode, defined as 
\begin{linenomath}
\begin{equation}
\label{eq:flux}
\Phi_{0}=\frac{({I-I_{\text{DCR}}})\lambda}{\text{Rhc}} 
\end{equation}
\end{linenomath}
where $I$ and $I_{\text{DCR}}$ are the Photo-Diode currents with and without illumination, respectively, R is the Photo-Diode responsivity at the wavelength $\lambda$ provided by NIST, h is Planck's constant and c is the speed of light. The PDE then follows as:
\begin{linenomath}
\begin{equation}
\label{eq:PDE_new}
\text{PDE}=\frac{\text{R}_{\text{0}}}{\Phi_{0}}
\end{equation}
\end{linenomath}
We emphasize that this technique is free from CAs since its contribution to the total pulse rate can be discriminated while constructing the rate plot, as shown in Fig.~\ref{fig:pulserateMEG2lightdark}.

\subsubsection{IHEP Technique}
\label{S:PDE_IHEP_tech}


The IHEP technique relies on the correlation between the SiPM current under $175$~nm light illumination and its PDE as follows: 
\begin{linenomath}
\begin{equation}
\label{eq:IHEP_PDE_174}
   \big(I_{\text{SiPM}}(V,175)-I^\text{DCR}_{\text{SiPM}}(V)\big)=\text{PDE}_{175}(V)\times\Phi_0(175)\times f(V)
\end{equation}
\end{linenomath}
where $V$ is the SiPM reverse bias voltage, $I_{\text{SiPM}}$ and $I^\text{DCR}_{\text{SiPM}}$ are the SiPM current with and without illumination, $\text{PDE}_{175}$ and $\Phi_0(175)$ are the SiPM PDE and the photon flux (Eq.~\ref{eq:flux}) measured at $175$~nm with the SiPM and the calibrated diode, respectively. All these quantities were measured by IHEP at 300~K. $f(V)$ is a correction factor that accounts for the SiPM gain and for the CA noise contribution that artificially increases the total current produced by the SiPM. It can be written as
\begin{equation}
    f(V)\sim q_e\times\big(1+\langle\Lambda\rangle\big)\times \overline{\text{G}}_{1\text{ PE}}
\end{equation}
Overall $f(V)$ is a function of the applied bias voltage $V$, but can be considered wavelength independent because it depends only on the SiPM intrinsic characteristics. More precisely the sensor gain $\overline{\text{G}}_{1\text{ PE}}$ depends on the SiPM SPAD capacitance while afterpulses and optical crosstalks that contribute to the sensor CAs $\langle\Lambda\rangle$ depend on impurities and cell geometry, respectively. In addition to its wavelength independence, $f(V)$  also shows a weak temperature dependence since the SiPM gain is almost temperature independent, as shown in Ref.~\cite{gallina2021development,Gallina2019} and the SiPM CAs has a weak temperature dependence, as previously discussed in Sec.~\ref{S:CA}. These fluctuations will not contribute significantly to the PDE measurements since the systematic error on the photon flux totally dominates Eq.~\ref{eq:IHEP_PDE_174}.\\\indent In this work, the correction factor $f(V)$ is estimated at 233~K using its wavelength and temperature independence by illuminating the SiPM with a pulsed visible light source (404~nm) as follows:
\begin{linenomath}
\begin{equation}
  f(V) = \frac{ \big(I_{\text{SiPM}}(V,404)-I^\text{DCR}_{\text{SiPM}}(V)\big)}{\upmu^{\text{SiPM}} \times f_L\times q_e}
\end{equation}
\end{linenomath}
where $I_{\text{SiPM}}$  and $I^\text{DCR}_{\text{SiPM}}$ are the SiPM current with and without 404~nm  illumination, $\upmu^{\text{SiPM}}$ is the average number of photons detected by the SiPM in each laser pulse of frequency $f_L$, and $q_e$ is the electron charge.  $\upmu^{\text{SiPM}}$  was measured by counting the number of laser flashes in which no pulses were detected ($N_{0}$). Using Poisson statistics, $\upmu^{\text{SiPM}}$ can be expressed as:
\begin{linenomath}
\begin{equation}
\upmu^{\text{SiPM}} = -\ln\left( \frac{N_{0}}{N_{\text{TOT}}}\right) - \upmu_{\text{DCR}}
\end{equation}
\end{linenomath}
where $N_{\text{TOT}}$ is the total number of laser pulses. This method is independent of CAs and it requires a correction for the average number of dark noise pulses in the acquisition window ($\upmu_{\text{DCR}}$).

\subsubsection{Photon Detection Efficiency at $175$~nm}
\label{S:PDE_174}

\begin{figure}[ht]
\centering
\includegraphics[width=0.99\linewidth]{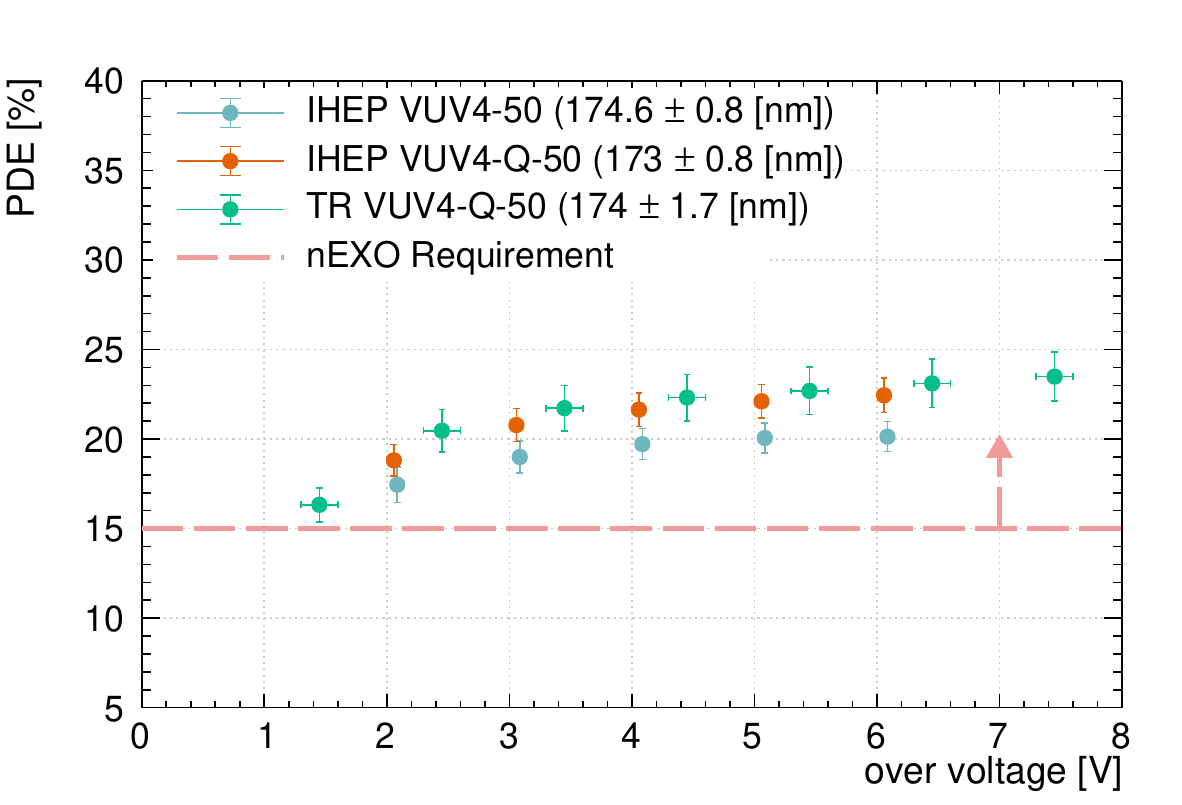}
\caption{Photon Detection Efficiency (PDE) measured at roughly~$~175$~nm as a function of the applied over voltage for HPK VUV4 MPPCs. The IHEP measurements are done at 300~K with a wavelength uncertainty of $\sim2$~nm FWHM. The TRIUMF ones instead at 163~K with a wavelength uncertainty of $\sim4$~nm FWHM. The error bars on each point  account for the presence both of the statistical and the systematic uncertainty. The dashed line represents the nEXO requirement.}
\label{fig:PDEVUV4s}
\end{figure}

\begin{figure}[ht]
\centering
\includegraphics[width=0.99\linewidth]{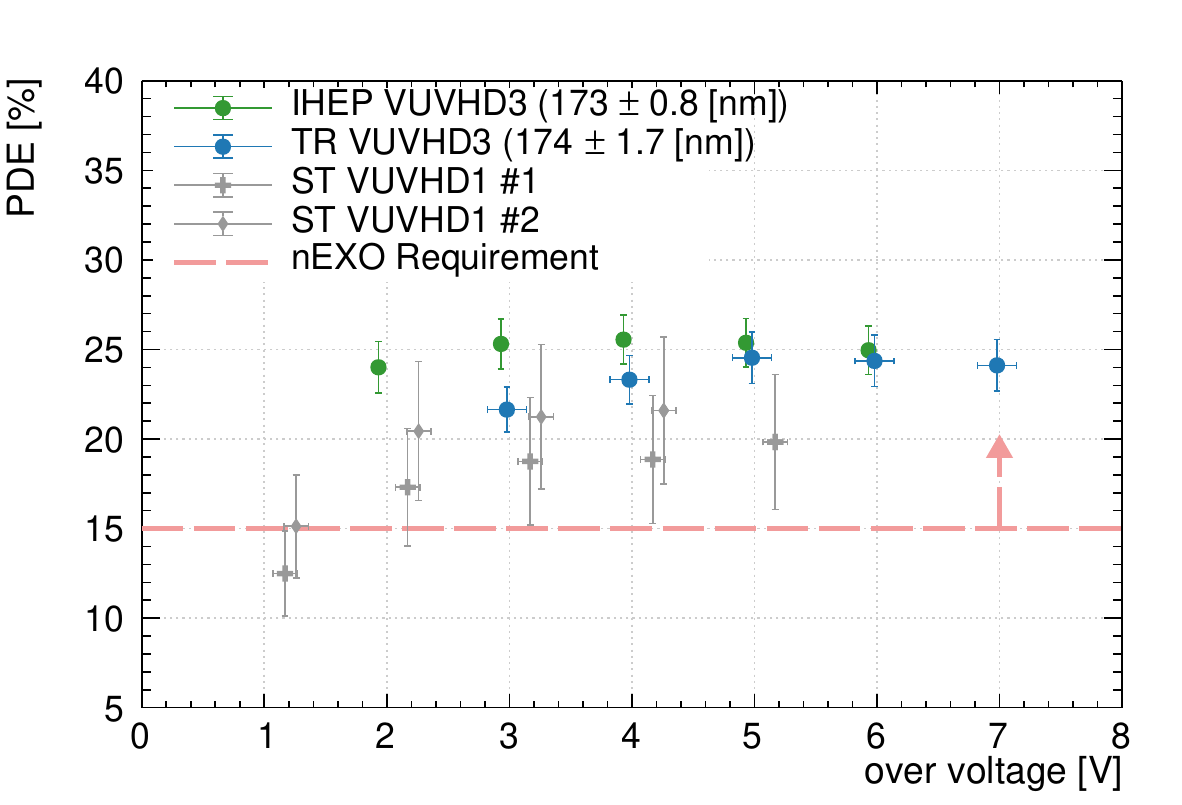}
\caption{Photon Detection Efficiency (PDE) measured at roughly $~175$~nm as a function of the applied over voltage for FBK VUVHD3 SiPMs. The IHEP measurements are done at 300~K with a wavelength uncertainty of $\sim2$~nm FWHM. The TRIUMF ones instead at 163~K with a wavelength uncertainty of $\sim4$~nm FWHM. The error bars on each point  account for the presence both of the statistical and the systematic uncertainty.  ST VUVHD1 \#1 and ST VUVHD1 \#2 are instead the PDE of FBK VUVHD1 SiPMs measured in Ref.~\cite{Ako}. The dashed line represents the nEXO requirement.}
\label{fig:PDEFBKs}
\end{figure}

Fig.~\ref{fig:PDEVUV4s} and Fig.~\ref{fig:PDEFBKs} show the PDE measured at $175$~nm and as a function of the applied over voltage for HPK VUV4 MPPCs and FBK VUVHD3 SiPMs, respectively. As already noted the IHEP measurements were done at 300~K while the TRIUMF ones at 163~K. The temperature dependence of the SiPM PDE is however expected to be weak, as mentioned in Sec.~\ref{S:PDE}. In Fig.~\ref{fig:PDEFBKs} we also report the PDE measured in Ref.~\cite{Ako} of the previous generation of FBK devices (FBK VUVHD1). Overall, the two generations of FBK SiPMs have an efficiency that is compatible, within uncertainties. However, it is worth noting that the new data have significantly smaller systematic uncertainties thanks to the NIST calibrated diodes used for these measurements. For instance, at an over voltage of 3~V we measure an average PDE of $20.5\pm1.1$\% for HPK MPPCs and of $24.3\pm1.4$\% for FBK VUVHD3 SiPMs. Both are well above the nEXO requirement.

\subsubsection{PDE wavelength dependence}
\label{S:PDE_wavelength}

\begin{figure}[ht]
\centering
\includegraphics[width=0.99\linewidth]{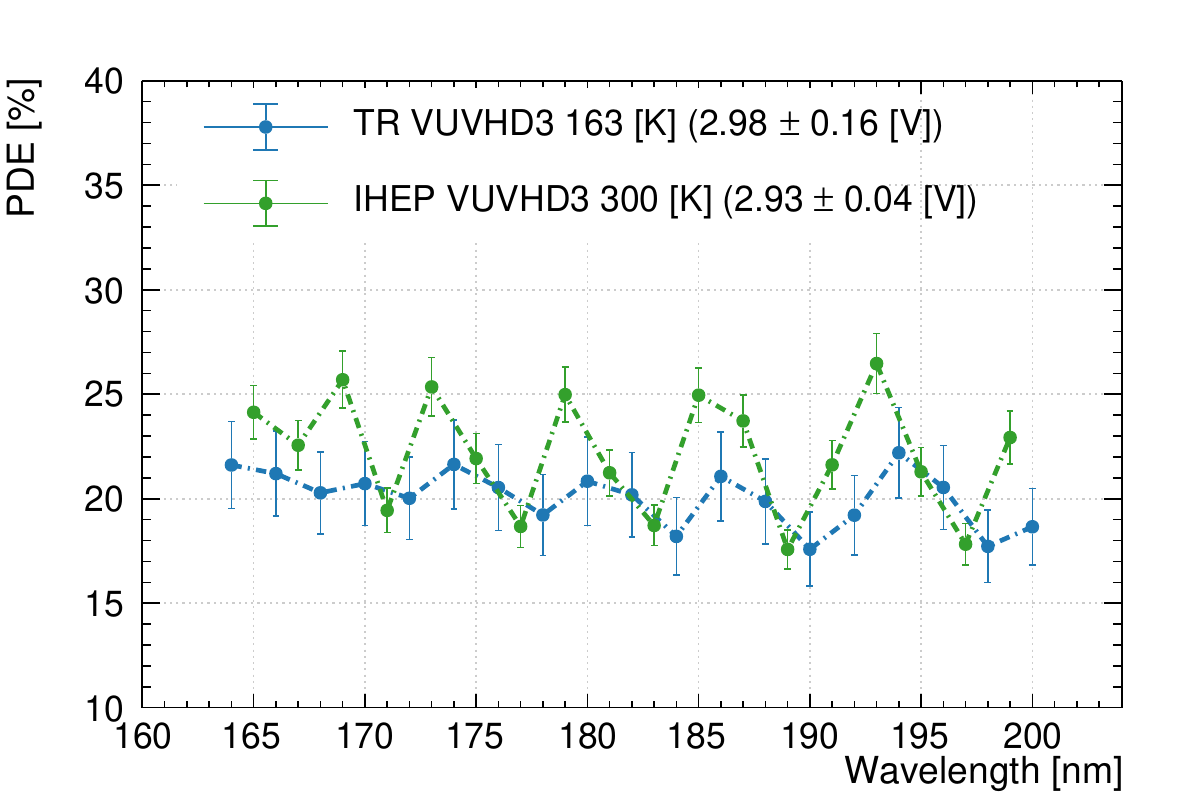}
\includegraphics[width=0.99\linewidth]{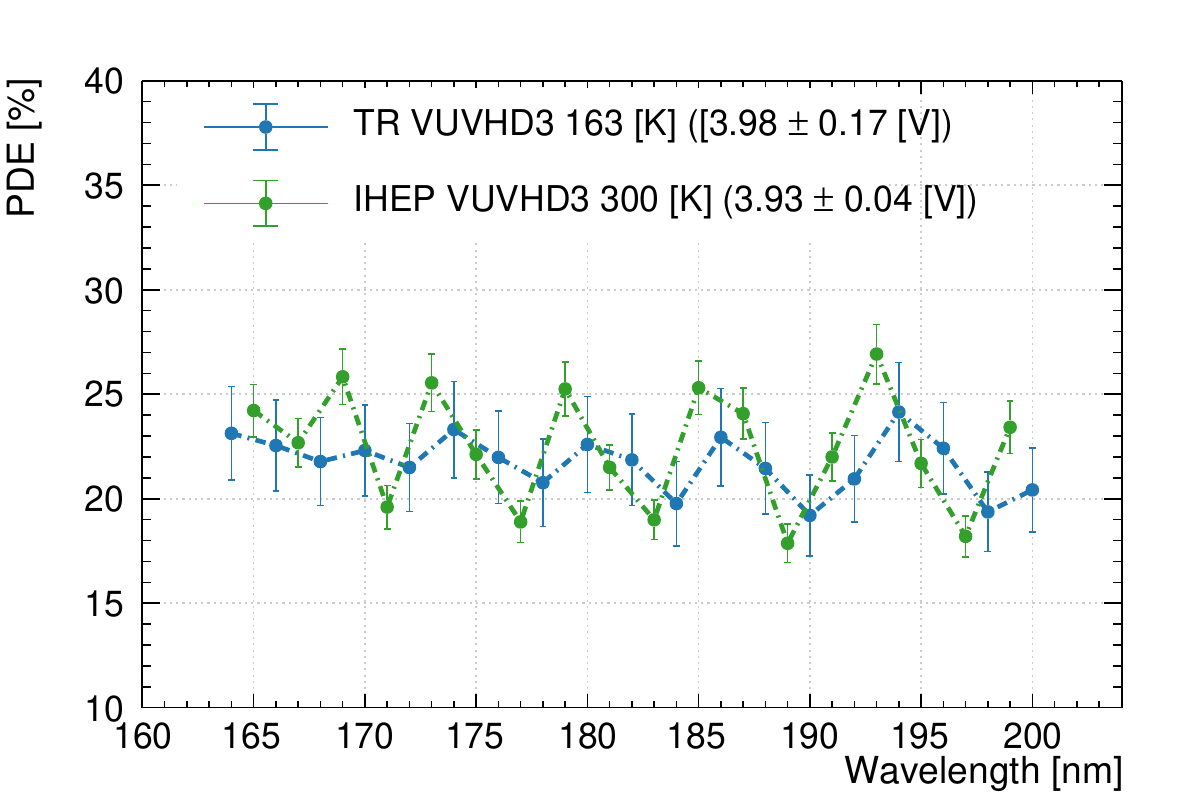}
\caption{Photon Detection Efficiency (PDE) measured as a function of the wavelength for 3~V and 4~V over voltage for FBK VUVHD3 SiPMs. The IHEP measurements are done at 300~K with a wavelength uncertainty of $\sim2$~nm FWHM. The TRIUMF ones instead at 163~K with a wavelength uncertainty of $\sim4$~nm FWHM. The error bars on each point account for the presence of the statistical and the systematic uncertainty.}
\label{fig:PDE_HD3_1}
\end{figure}

\begin{figure}[ht]
\centering
\includegraphics[width=0.99\linewidth]{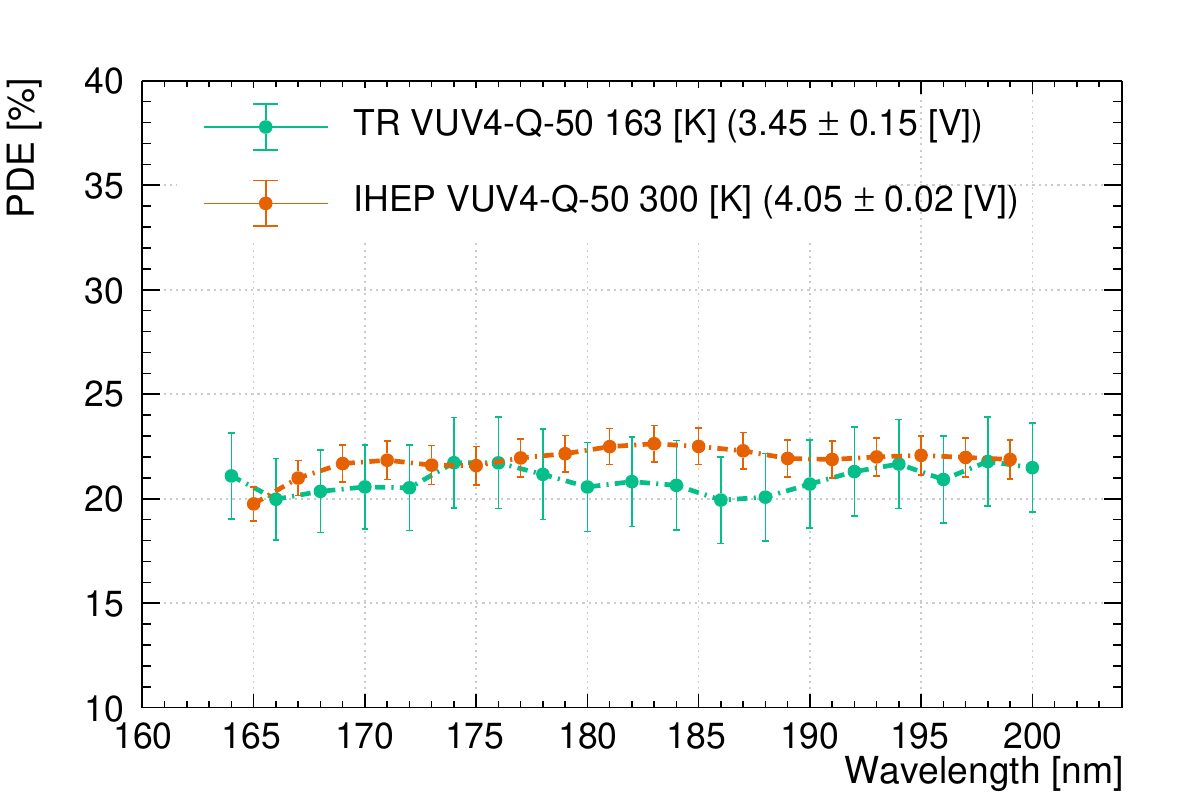}
\includegraphics[width=0.99\linewidth]{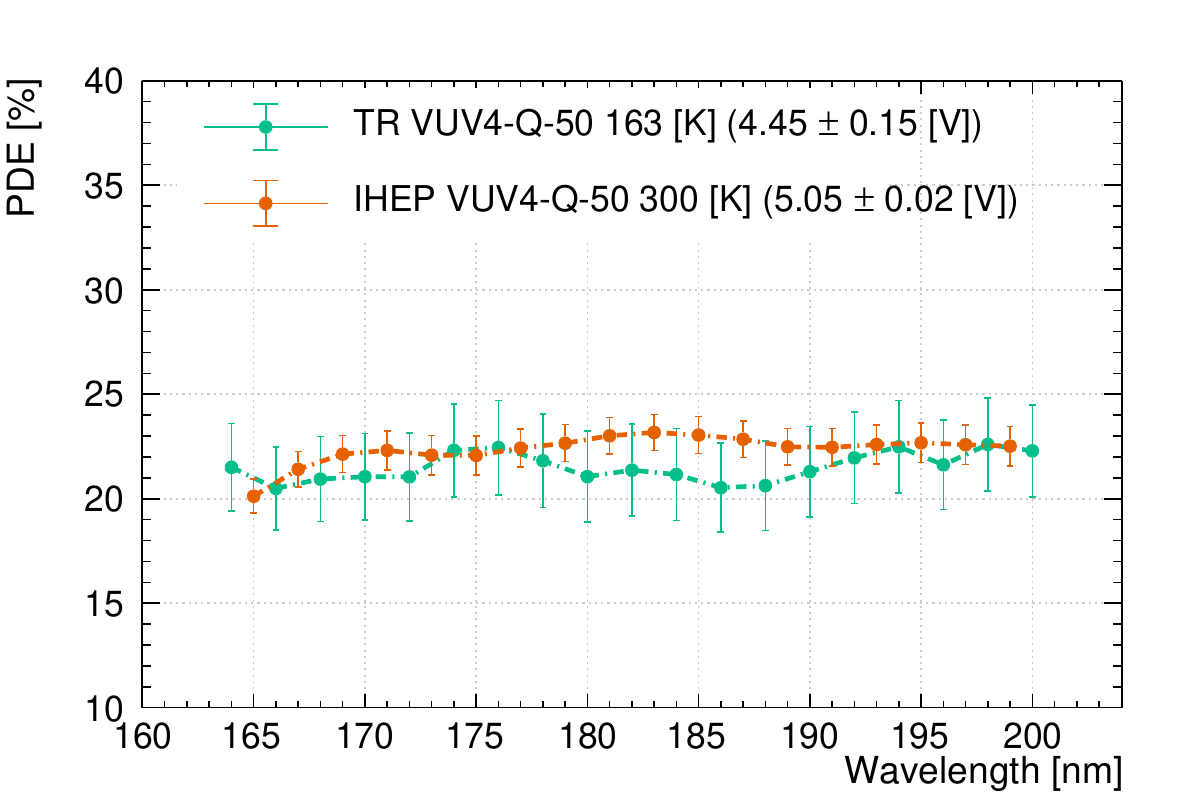}
\caption{Photon Detection Efficiency (PDE) measured as a function of the wavelength for roughly 4~V and 5~V of over voltage for HPK VUV4-Q-50 MPPCs. The IHEP measurements are done at 300~K with a wavelength uncertainty of $\sim2$~nm FWHM. The TRIUMF ones instead at 163~K with a wavelength uncertainty of $\sim4$~nm FWHM.  The error bars on each point account for the presence of the statistical and the systematic uncertainty.}
\label{fig:PDE_VUV4_2}
\end{figure}

The LXe scintillation emission spectrum is in the shape of a Gaussian function, with the mean at $174.8~\text{nm}$ and a FWHM of  $10.2~\text{nm}$ \cite{FUJII2015293}. For precise detector simulation it is therefore necessary to know not only the PDE at the maximum of the emission spectrum, but also in a broader wavelength range. In Sec.~\ref{S:PDE_174} we have shown the $175$~nm PDE measured by TRIUMF and IHEP as a function of the applied over voltage at 163~K and 300~K for all the devices under test. The wavelength dependence of the SiPM PDE can be extracted using the PDE measured at $175$~nm and by utilizing a correlation between the SiPM current under illumination and its PDE, similar to the one used by IHEP in Sec.~\ref{S:PDE_IHEP_tech}. More precisely, if the SiPM PDE is known at a specific wavelength ($175$~nm in this case) and for a specific reverse bias voltage $V$, Eq.~\ref{eq:IHEP_PDE_174} can be used to estimate the wavelength independent correction factor $f(V)$ and to measure the SiPM PDE for the same bias voltage $V$, but for a different wavelength $\lambda$, as follows:
\begin{linenomath}
\begin{equation}
\text{PDE}_{\lambda}(V)=\frac{\big(I_{\text{SiPM}}(V,\lambda)-I^\text{DCR}_{\text{SiPM}}(V)\big)}{\Phi_0(\lambda)\times f(V)}
\end{equation}
\end{linenomath}
where $\Phi_{0}(\lambda)$ is defined in Eq.~\ref{eq:flux},  
$I_{\text{SiPM}}$ and $I^\text{DCR}_{\text{SiPM}}$ are the SiPM current with and without the $\lambda$ illumination.\\\indent As an example in Fig.~\ref{fig:PDE_HD3_1} we report the PDE of the FBK VUVHD3 SiPMs measured at roughly 3 and 4~V of over voltage by TRIUMF and IHEP in the wavelength range 165-200 nm at 163~K and 300~K, respectively. Similar figures hold for other over voltages. The error bars on each point include statistical and systematic uncertainties. The FBK PDE data show clear oscillations due to interference of the incident light in the $\sim1.5\upmu\text{m}$ thick $\text{SiO}_2$ cover layer deposited on the surface of these SiPMs.
A similar interference pattern was also seen in a previously reported nEXO study where the specular reflectivity of FBK VUVHD1 SiPMs was measured as a function of incidence angle and wavelength \cite{Lv2020}. The FBK VUVHD1 and FBK VUVHD3 indeed share the same surface coating configuration. Oscillations are expected to be damped in LXe due to an excellent match between the LXe and $\text{SiO}_2$ index of refraction~\cite{Lv2020}. Overall, the TRIUMF and IHEP measurements give comparable results within errors with well aligned maxima and minima of the interference pattern. The IHEP data, however, show a slightly larger peak-to-valley ratio as compared to the TRIUMF data, attributed to better wavelength resolution.\\\indent Fig.~\ref{fig:PDE_VUV4_2} shows the PDE of HPK VUV4-Q-50 MPPCs measured at roughly 4 and 5~V of over voltage by TRIUMF and IHEP in the wavelength range 165-200 nm and at 163~K and 300~K, respectively. Similar figures hold for other over voltages. Again, error bars on each point include statistical and systematic uncertainties. Unlike FBK SiPMs, the HPK VUV4-Q-50 MPPCs do not exhibit an interference pattern in their PDE, most likely because their surface coating is thinner\footnote{HPK didn't disclose the configuration and chemistry of the surface coating structure. However 
accordingly to the HPK documentation, HPK VUV4-50 and HPK VUV4-Q-50 MPPCs should share the same, unknown, surface coating topology.}. The absence of an interference pattern is compatible with the specular reflectivity measurements reported in Ref.~\cite{Lv2020}. 

\begin{figure}[ht]
\centering\includegraphics[width=0.99\linewidth]{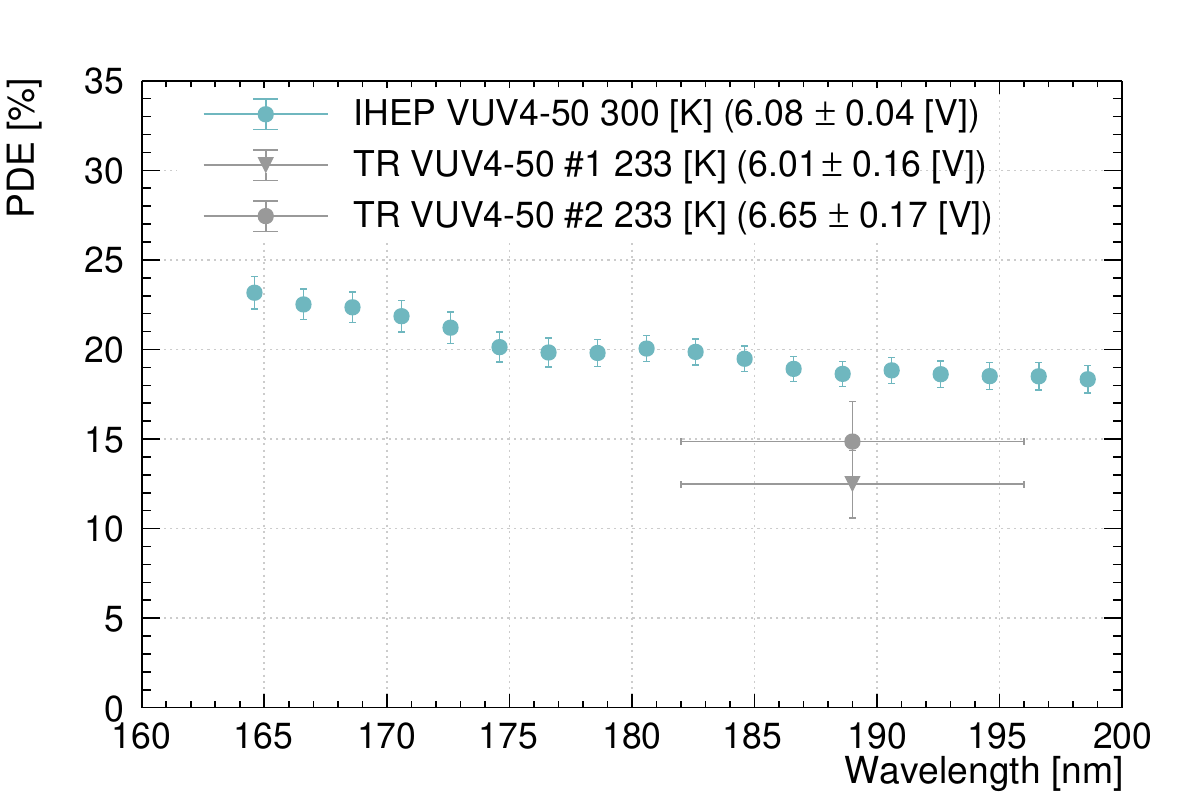}
\caption{Photon Detection Efficiency (PDE) measured by IHEP as a function of the wavelength for roughly 6~V of over voltage for HPK VUV4-50 MPPCs. For reference in the same figure we also reported the PDE measured in Ref.~\cite{Gallina2019} for the nominally same type of MPPC, but different test sample. Due to the different filtering scheme the wavelength uncertainty of the previously published measurement was significantly larger.}
\label{fig:PDE_single}
\end{figure}

Fig.~\ref{fig:PDE_single} shows instead the PDE of HPK VUV4-50 MPPC measured by IHEP at roughly 6~V of over voltage in the wavelength range 165-200 nm at 300~K. In the same figure we also reported the PDE measured in Ref.~\cite{Gallina2019} at roughly the same over voltage for nominally the same type of MPPC (HPK VUV4-50 MPPC), but different test sample. The comparison in Fig.~\ref{fig:PDE_single} is done at 6~V of over voltage to not be directly sensitive to the applied over voltage since at large over voltages the PDE of both photosensors is, within uncertainties, independent on the applied over voltage. The new measurements confirm that this type of MPPC (HPK VUV4-50) has a PDE that is below the almost flat 20-25\% advertised by HPK in the same wavelength range \cite{Doc_VUV4}. Moreover, the HPK VUV4-50 has a lower PDE than the corresponding quad device (HPK VUV4-Q-50), as shown in Fig.~\ref{fig:PDE_VUV4_2}. The new IHEP measurements show a slightly larger PDE than previously reported, likely due to device to device non-uniformity. Indeed, the previously reported results already showed a large spread in the PDE, as shown in Fig.~\ref{fig:PDE_single}.

\section{Estimation of the nEXO Energy Resolution}
\label{S:energyresolution}

The nEXO experiment is designed with optimized scintillation light and charge collection to provide an excellent energy resolution, with a final design goal of $\sigma/Q<1\%$ for the 0$\nu \beta \beta$ decay of \ce{^{136}Xe} ($Q=2458.07\pm0.31~\text{keV}$ \cite{Redshaw2007,PhysRevC.82.024603}). Recently, other LXe TPCs with high light collection efficiency, such as Xenon1T and LZ have featured sub-percent detector energy resolution over the relevant energy range~\cite{Aprile,Pereira}.\\Important differences between nEXO and these dual phase TPCs include the use of aluminised surfaces instead of PTFE as light reflectors, the direct detection of charge without amplification, and the positioning of light sensors on the detector barrel, behind an optically open field-shaping electrode cage~\cite{NEXOCollaboration2018,nEXO:2020mec}\footnote{PTFE is a better (diffuse) reflector than aluminum, as shown in Ref.~\cite{Neves2017}. In contrast, the planned aluminised surfaces (Al+$\text{MgF}_{2}$) only arrives up to 80-90\%. Even though the reflectivity is lower, this surface coating is chosen for compatibility with the open field cage design where light can be incident on the metal field cage components themselves.}. \\\indent In Ref.~\cite{Ako} we have shown that the first generation of FBK devices (FBK~VUVHD1) satisfy the nEXO requirement with an optimal energy resolution achieved using an over voltage between 2~V and 3~V. In this section we evaluate the energy resolution achievable by nEXO using the better-performing VUV sensitive SiPMs tested in this work.\\\indent The production of scintillation light in xenon proceeds via a few paths \cite{Doke:2002oab}. In general, when a particle deposits an energy $E$ in LXe, it produces heat, electron-ion pairs and atomic excitations (forming excitons when in liquid). Excitons promptly form excited dimers, {\it i.e.} a Xe$^*_2$ molecule, while ions  form charged dimers \cite{Martin1971}. Excited xenon dimers decay to the ground state producing 175 nm scintillation light. The charged dimers, instead,  neutralize by capturing a free electron. This recombination process is dissociative and results in an excited xenon atom which again forms an excited dimer similarly to the direct excitation channel mentioned above \cite{PhysRevB.20.3486}.\\Other processes may occur, such as quenching and intermediate transitions, or atomic relaxation~\cite{PhysRevB.27.5279,wasterdalethesis}.\\\indent The energy resolution model derived in this section represents a progression of the EXO-200 semi-empirical one presented in Ref.~\cite{Collaboration2019}, and it is  based on the assumption that each recombining electron-ion pair always produces an exciton which in turn produces a photon. Fits to the EXO-200 detector response in Ref.~\cite{Collaboration2019} are consistent with this assumption, indicating that at most a few percent of electron-ion pairs might not produce a scintillation photon upon recombination. 
If we denote with $r$ the fraction of recombining electrons, the maximum number of detectable electrons ($n_q$) and photons ($n_p$) for an initial population of $n_i$ electron-ion pairs and $n_\text{ex}$ excitons, are $n_q=(1-r)n_i$ and $n_p=(n_\text{ex}+r n_i)$, respectively. Under these assumptions, it is possible to define a recombination-independent value $W\equiv E/(n_q+n_p)=E/(n_i+n_\text{ex})$, which corresponds to the energy required to create a single quantum of either type (light or charge).\\\indent Because the light and the charge channels are anti-correlated, the mean number of quanta (of both types) produced by a single energy deposition can be written as~\cite{Collaboration2019} 
\begin{linenomath}
\begin{linenomath}
\begin{equation}
\label{eq:quanta_number}
 \langle n\rangle = \frac{\langle E\rangle}{W}\propto\cos(\theta)\langle n_q\rangle  + \sin(\theta)\langle n_p\rangle
\end{equation}
\end{linenomath}
\end{linenomath}
where, for nEXO, $\langle E\rangle$ is the $Q$-value of the $0\nu \beta \beta$ decay of $\ce{^{136}Xe}$ and $W=11.5\pm 0.5~\text{(syst.)}\pm0.1~\text{(stat.)}~\text{eV}$, as measured by EXO-200~\cite{Collaboration2019}. This value of $W$ is smaller than what is currently used in the NEST code~\cite{Szydagis2011}, but is confirmed by recently published results~\cite{Baudis2021_2}.
$\theta$ is the rotation angle in the charge-light 2-dimensional space and indicates the optimal weighting of the two signals that minimizes the energy resolution \cite{NEXOCollaboration2018}. \\\indent The relative standard deviation associated with the quanta counting of Eq.~\ref{eq:quanta_number} (also called the energy resolution for the $0\nu \beta \beta$ decay) is computed assuming that the optimal rotation angle is $\theta = \pi/4$~\cite{Collaboration2019}: 
\begin{linenomath}
\begin{equation}
\label{eq:energy_res}
    \frac{\sigma_n}{ \langle n\rangle}=\frac{\sqrt{\sigma^2_q+\sigma^2_p+2\text{Cov}_{q,p}+\sigma^2_\text{Xe}}}{\langle n\rangle}
\end{equation}
\end{linenomath}
where $\sigma^2_q$ and $\sigma^2_p$ are the variances of $n_q$ and $n_p$, respectively and $\text{Cov}_{q,p}$ is their covariance. The additional term in Eq.~\ref{eq:energy_res} is a variance parameterised by a Fano factor-like term: $\sigma^2_\text{Xe}=f_\text{Xe}\,\langle n\rangle$ that accounts for intrinsic fluctuations in the total number of quanta unrelated to recombination ({\it e.g.}, energy loss to heat) \cite{Collaboration2019}. While never been measured for LXe, $f_\text{Xe}$ has been calculated to be 0.059 \cite{RevModPhys.82.2053,Seguinot1995} and is poorly constrained by data since it is a sub-dominant contribution to the resolution of existing detectors. Compared to the charge readout noise and photon collection fluctuations, this additional term also represents a subdominant contribution for nEXO and will be neglected in the following sections.

\subsection{Contribution of the scintillation detector}
\label{S:lightchannel}

The contribution of the light subsystem performance to the total nEXO energy resolution can be derived by considering two assumptions. First, the number of photons detected ($n_d$) follows a binomial distribution~\cite{Collaboration2019} with detection probability equal to
\begin{linenomath}
\begin{equation}
\label{eq:epsilon_p}
    \epsilon_p=\text{PTE}\times\frac{\text{PDE}}{1-\text{R}}
\end{equation}
\end{linenomath}
where PTE and PDE are the Photon Transport and Detection Efficiency for Xe scintillation light\footnote{The SiPM PDE is divided by the SiPM transmission $\text{T}=(1-\text{R})$ to avoid to double count this term in the PTE contribution.} ($\sigma_{d}^2=\epsilon_p(1-\epsilon_p)n_p $, $\langle n_d\rangle=n_p\epsilon_p$). Second, the number of Dark Count (DC) events ($n_{DC}$) in the acquisition window follows a Poisson distribution.\\\indent The fluctuation in the number of measured photons ($\sigma^2_p$) can then be written as 
\begin{linenomath}
\begin{equation}
\begin{aligned}
\label{eq:light_noise_2}
\sigma^2_p&=\frac{(1-\epsilon_p)n_p}{\epsilon_p}+\frac{\langle n_{DC}\rangle+\langle n_d\rangle}{\epsilon_p^2}\times\frac{\sigma^2_{\Lambda}}{(1+\langle\Lambda\rangle)^2}+\\
&+\frac{\langle n_{DC}\rangle}{\epsilon_p^2}+n^2_p\sigma^2_{lm}+
\frac{\eta^2_{noise}}{\epsilon_p^2(1+\langle\Lambda\rangle)^2}+\sigma^2_r
\end{aligned}
\end{equation}
\end{linenomath}
where: (i) $\langle\Lambda\rangle$ is the average extra charge produced by CAs per pulse, (ii) $\sigma_{\Lambda}$ is its standard deviation, (iii) $\sigma^2_r$ accounts for the fluctuation in units of quanta, resulting from recombination (iv) $\sigma^2_{lm}$ 
accounts for some residual calibration uncertainty, which systematically biases $\langle n_p \rangle$ ({\it e.g.} due to spatial variations that cannot be completely calibrated), and (v) $\eta^2_{noise}$ accounts for an additional noise contribution due to distortions by the electronic noise in the photon-readout channel. This quantity cannot be calibrated until the finalisation of the nEXO photon-readout, but based on the requirement reported in Sec.~\ref{S:intro} ($<0.1$~PE~r.m.s.), it is expected to be a subdominant contribution to the energy resolution. The mean number of dark noise events in Eq.~\ref{eq:light_noise_2} is also expected to be a sub-dominant contribution, as shown by Fig.~\ref{F:DN_OV}. For these reasons both quantities will henceforth be neglected.\\Eq.~\ref{eq:light_noise_2} can then be simplified as
\begin{linenomath}
\begin{equation}
\begin{aligned}
\label{eq:light_noise}
\sigma^2_p\sim\frac{(1-\epsilon_p)n_p}{\epsilon_p}+\frac{\langle n_d\rangle}{\epsilon_p^2}\times\frac{\sigma^2_{\Lambda}}{(1+\langle\Lambda\rangle)^2}+n^2_p\sigma^2_{lm}+\sigma^2_r
\end{aligned}
\end{equation}
\end{linenomath}
Eq.~\ref{eq:light_noise} shows how the SiPM CAs contribute to the light component of the energy resolution with a term proportional to
$\sigma_{\Lambda}/(1+\langle\Lambda\rangle)$, providing a physics-driven motivation for the requirement introduced in Sec.~\ref{S:CA}.\\Generally, the energy resolution worsens at high over voltages since $\sigma_{\Lambda}$ increases faster than~$\langle\Lambda\rangle$. \\\indent In addition to CAs and dark noise, SiPMs are also affected by external crosstalk, introduced in Sec.~\ref{S:CDP} and not explicitly included in Eq.~\ref{eq:light_noise}.
The number of external crosstalk photons emitted per avalanche depends on the SiPM gain. Moreover the infrared light is the main component of the SiPM secondary photon emission, as shown in Ref.~\cite{McLaughlin2021}. Preliminary studies show that SiPMs secondary photon emission could degrade the nEXO energy resolution at high over voltages, while suggesting a subdominant contribution at low over voltages ($\le3$~V)~\cite{Boulay2022}. Its impact on the energy resolution, however, depends on the nEXO TPC PTE for infrared light that, in turn, depends on the reflectivity of the TPC materials and the SiPM PDE at these wavelengths. Both quantities are so-far not well known for infrared wavelengths and await new measurements necessary to fully predict their impact on the nEXO energy resolution.

\subsection{Contribution of the ionization detector}
The contribution of the charge subsystem to the nEXO energy resolution is derived similarly to what is presented in Sec.~\ref{S:lightchannel}. The number of charges detected ($n_{dq}$) can be written as
\begin{linenomath}
\begin{equation}
n_{dq}=\epsilon_qn_q
\end{equation}
\end{linenomath}
with $ \epsilon_q=e^{-t/\tau}$ the charge collection efficiency, a function of the mean drift time $t$ in LXe and of the electron lifetime $\tau$, a finite quantity which depends on the concentration of electronegative contaminants. The drift time $t$ is derived from the drift length $l$ and velocity $v$ as $l/(2v)$ where the factor 2 accounts for the averaging of signal over the entire fiducial volume. The contribution of the charge subsystem to the total nEXO energy resolution, $\sigma_q^2$, can then be derived assuming binomial statistics~\cite{Collaboration2019} for the number of detected charges with probability of detection equal to $\epsilon_q$, as follows
\begin{linenomath}
\begin{equation}
\label{eq:charge_noise}
\sigma_q^2=\frac{n_{q}t}{\tau}+\frac{\sigma^2_{q,noise}}{\epsilon_q^2}+\sigma_r^2
\end{equation}
\end{linenomath}
where $\sigma^2_{q,noise}$ accounts for the electronic noise contribution in the charge channel and $\sigma_r^2$ is the variation in units of quanta in the charge channel due to recombination fluctuations. Possible deviations from the binomial statistics may be related to the change in charge loss versus drift length. A drift time correction will be applied to remove the average charge loss on an event-by-event basis. Moreover here we are ignoring any residual calibration uncertainty in the correction. However, statistical fluctuations in the charge loss for a given event can still contribute to the resolution. For the $>5$~ms electron lifetime in nEXO, these fluctuations will not contribute significantly to the charge resolution since electronics noise dominates Eq.~\ref{eq:charge_noise}.

\subsection{Predicted Energy Resolution}
\begin{figure}[ht]
\centering\includegraphics[width=0.99\linewidth]{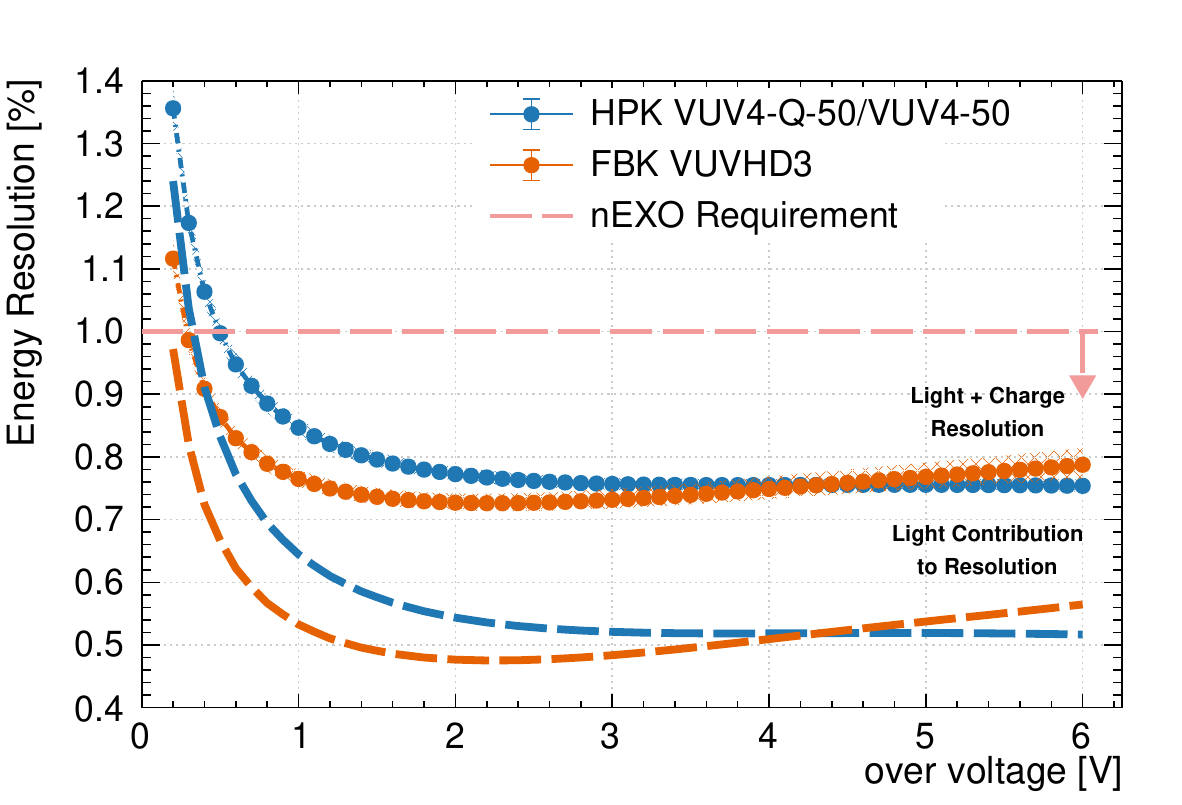}
\caption{Estimated energy resolution ($\sigma_n/\langle n\rangle$, Eq.~\ref{eq:energy_res_final}) as a function of the applied over voltage for the 0$\nu \beta \beta$ decay of \ce{^{136}Xe} ($Q=2458.07\pm 0.31~\text{keV}$) achievable by the nEXO detector with the VUV-sensitive SiPMs tested in this work (FBK VUVHD3 SiPMs and HPK VUV4 MPPCs). Dashed lines represent the contribution of the light channel ($\sigma_p/\langle n\rangle$) to the total energy resolution, neglecting recombination fluctuations ($\sigma^2_r$). The LXe light yield is derived from Ref.~\cite{Collaboration2019} for a drift field of  400~V/cm, the nominal one in the actual nEXO design. Uncertainty bands on the light channel-only resolution are not shown for clarity.}
\label{F:energy_resolution}
\end{figure}

The final predicted nEXO energy resolution is obtained by substituting the contribution of the light (Eq.~\ref{eq:light_noise}) and charge (Eq.~\ref{eq:charge_noise}) channels in Eq.~\ref{eq:energy_res}, and assuming perfect anti-correlation between the two channels ($\text{Cov}_{q,p}=-\sigma_r^2$), as follows:
\begin{linenomath}
\begin{equation}
\begin{aligned}
   \frac{\sigma_n}{ \langle n\rangle}&=\frac{\sqrt{\Big(\frac{(1-\epsilon_p)n_p}{\epsilon_p}+\frac{n_p}{\epsilon_p}\cdot\frac{\sigma_{\Lambda}^2}{(1+\langle\Lambda\rangle)^2}+n^2_p\sigma^2_{lm}\Big)+\Big(\frac{n_{q}t}{\tau}+\frac{\sigma^2_{q,noise}}{\epsilon^2_q}\Big)}}{\langle n\rangle}
\end{aligned}
\label{eq:energy_res_final}
\end{equation}
\end{linenomath}
Fig.~\ref{F:energy_resolution} shows the predicted energy resolution for the 0$\nu \beta \beta$ decay $Q$-value of \ce{^{136}Xe} as a function of the photosensors over voltage and Table~\ref{T:energy_res_table} summarises the parameters used for the calculation. The SiPM CAF is computed as in Sec.~\ref{S:CA} (Fig.~\ref{fig:ratio}). The SiPMs and MPPCs PDE are computed by using a polynomial spline interpolation (forced to go to zero at 0~V of over voltage) of all the data of Figs.~\ref{fig:PDEVUV4s},\ref{fig:PDEFBKs}. The SiPM reflectivity in Eq.~\ref{eq:epsilon_p} is derived from the almost normal incidence 175~nm vacuum data ($\sim5^{\circ}$) published in Ref.~\cite{Lv2020}, under the assumption of unpolarised light\footnote{Ref.~\cite{Lv2020} reports the specular reflectivity of FBK VUVHD1 SiPMs and HPK VUV4-50 MPPCs as a function of the incident wavelength and angle.  As noted in Sec.~\ref{S:PDE_wavelength}, the FBK VUVHD1 and VUVHD3 SiPMs share the same surface coating structure. Although HPK didn’t disclose the surface properties of its devices, its documentation indicates that HPK VUV4-50 and HPK VUV4-Q-50 MPPCs have the same surface coating.}. The LXe light yield is instead derived from Ref.~\cite{Collaboration2019} for a drift field of  400~V/cm, the nominal one in the actual nEXO design. The detector PTE, residual calibration uncertainty, noise in the charge channel, electron drift velocity and lifetime are taken from Ref.~\cite{Adhikari2022}. 
\begin{table}[ht]
\setlength\tabcolsep{5pt} 
\centering
\begin{tabular}{cccc}
   \toprule
   Symbol &  Meaning & Value  & Ref.\\
   \midrule     
    $Q$ [keV] & $Q$-value & 2458.07 &  \cite{Redshaw2007,PhysRevC.82.024603}\\
    W [$\text{eV}$] & Energy for 1 quantum  & 11.5 & \cite{Collaboration2019,Baudis2021_2}\\
    $n=Q/\text{W}$ [\#] & Number of quanta & 213745 & -\\
     $\gamma_p$ [$\gamma/\text{eV}$] & Light yield  & 0.037 & \cite{Collaboration2019}\\
    $ n_p=Q\times \gamma_p$ [$\#$] &     Number of photons &      90949 & -\\
  PTE [\#] &   Photon Transport Eff.  &     33.3\%  & \cite{Adhikari2022}\\
  $\sigma_{lm}$ [\#] &    Res. calib. uncertainty  &     0.5\% & \cite{Adhikari2022}\\
  PDE [\#] &    Photon Detection Eff.  &  Sec.~\ref{S:PDE} & -\\
  R [\#] &  Reflectivity \makecell{FBK\\HPK}  & \makecell{$27.7\pm1.6$\% \\$20\pm1$\%}   & \cite{Lv2020} \\
    $\epsilon_p$ [\#] & Total Photon Det. Eff. & Eq.~\ref{eq:epsilon_p} & -\\
  $\langle\Lambda\rangle$ [PE] &    Mean of CA &    Sec.~\ref{S:CA} & -\\
   $\sigma_{\Lambda}$ [PE] &     RMS of CA    & Sec.~\ref{S:CA} & -\\
    $ n_q=n-n_p$ [$\text{e}^-$] &     Number of electrons &   122797    & -\\
   $ \sigma_{q,noise}$ [$\text{e}^-$] &     Noise charge ch.  &      1132 & \cite{Adhikari2022}\\
   $l$ [m]  &    Drift length   &   1.187  & \cite{Adhikari2022}\\
    $v$ [$\text{mm}/\upmu\text{s}$]  &    Drift velocity &   1.73   & \cite{Adhikari2022} \\
    $\tau$ [ms]  &   Electron  lifetime  &   10  & \cite{Adhikari2022}\\
      $\epsilon_q=e^{-t/\tau}$ [\#] &    Charge coll. eff. &     96.6\%  & -\\
   \bottomrule
\end{tabular} 
\caption{Summary of the parameters of Eq.~\ref{eq:energy_res_final}. The SiPM Correlated Avalanches (CAs), RMS and Photon Detection Efficiency (PDE) are derived by interpolating all the data reported in Sec.~\ref{S:CA} and Sec.~\ref{S:PDE}. The SiPM reflectivity is derived from the almost normal incidence 175~nm vacuum data previously published in Ref.~\cite{Lv2020}, under the assumption of unpolarised light. The detector Photon Transport Efficiency (PTE), residual calibration uncertainty, noise in the charge channel, electron drift velocity, lifetime and light yield are extrapolated from Refs.~\cite{Adhikari2022,Collaboration2019} under the assumption of a drift field of 400~V/cm, the nominal one in the actual nEXO design. 
}
\label{T:energy_res_table}
\end{table}
The dashed lines in Fig.~\ref{F:energy_resolution} represent the contribution of the light channel ($\sigma_p/\langle n\rangle$) to the total energy resolution, obtained using Eq.~\ref{eq:light_noise} and neglecting the recombination fluctuation term ($\sigma^2_r$). Additionally, Fig.~\ref{F:energy_resolution} shows also the nEXO design specification.
Overall FBK VUVHD3 SiPMs and HPK VUV4 MPPCs are both excellent candidates for the nEXO light detection subsystem, each not only satisfying, but exceeding, the 1\% energy resolution requirement. This is a remarkable improvement as compared to the SiPMs tested in Refs.~\cite{Ako,Gallina2019}, in particular for previous generation HPK MPPCs that only marginally met the SiPM PDE requirement (Sec.~\ref{S:PDE}). The FBK VUVHD3 SiPMs have a slightly better energy resolution, especially at low over voltage ($\le$3~V), due to their higher PDE. HPK MPPCs, on the other hand, show a remarkably small degradation of the energy resolution with increased over voltage due to lower average extra charge produced by CAs per pulse, as shown in Sec.~\ref{S:CA}.\\\indent For instance, at 3~V of over voltage, we predict an energy resolution at the $^{136}$Xe decay $Q$-value of $0.73\pm0.02$\% for the FBK VUVHD3 SiPMs and of $0.76\pm0.01$\% for the HPK VUV4 MPPCs. These values are close to the $\sim0.8$\% extrapolated with the full nEXO Monte Carlo (MC) simulation~\cite{Adhikari2022}. The small discrepancy is mainly due to the conservative SiPMs PDE value, derived from earlier measurements of HPK devices~\cite{Gallina2019,Ako}, coded in the MC compared to the one, $5-10$\% higher, presented in this work. It should be stressed that the uncertainty on the energy resolution presented here only includes systematic uncertainties defined by the measurement reported in this paper. Other systematic effects related, {\it e.g.}, to noise on the charge channel would also impact the energy resolution.

\section{Conclusions}

\begin{table}[ht]
\centering
\begin{tabular}{ccc}
   \toprule
   Quantity & FBK VUVHD3 & HPK MPPCs \\
   \midrule
   DCR [Hz/mm$^{2}$] & $0.19\pm0.01$ & $0.35\pm 0.01$ \\
   $\langle\Lambda\rangle$ [PE] & $0.23\pm0.06$ & $0.06\pm0.02$ \\
   $\sigma_{\Lambda}$ [PE] &  $0.51\pm0.06$ & $0.25\pm0.01$\\
   CAF (Eq.~\ref{eq:ratio_CA}) [\#] & $0.42\pm0.07$ & $0.24\pm0.02$\\
   $\text{PDE}_{\text{175~nm}}$ [\#] & $24.3\pm1.4$\% & $20.5\pm1.1$\% \\
  Energy Resolution [\#] & $0.73\pm0.02$\% & $0.76\pm0.01$\%\\
   \bottomrule
\end{tabular} 
\caption{
Summary of the results derived for the characterization of the FBK VUVHD3 SiPMs and HPK VUV4 MPCCs and useful for nEXO operations. The dark count rate (DCR), the average extra charge produced by correlated avalanches per pulse in the $1\upmu\text{s}$ following the trigger pulse $\langle\Lambda\rangle$, its RMS $\sigma_{\Lambda}$, and the corresponding Correlated Avalanche Fluctuation (CAF), as defined in Eq.~\ref{eq:ratio_CA}, are reported for a temperature of 163~K and at an over voltage of 3~V. The Photon Detection Efficiency (PDE) is also reported for an over voltage of $3$~V, at 163~K, and for a mean wavelength of $175$~nm.}
\label{T:summary}
\end{table}

This paper describes measurements performed by the nEXO collaboration to characterize the properties of new VUV sensitive SiPMs at 163~K. In particular, this work focused on FBK VUVHD3 SiPMs and HPK VUV4 MPPCs, identified as possible options for the nEXO experiment. The results of the characterization that are relevant for the nEXO detector are summarized in Table~\ref{T:summary}. For  a  device  temperature  of  163~K and at an over voltage of 3~V,  the  dark  noise  rates of FBK~VUVHD3 SiPMs and HPK VUV4 MPPCs are measured to be $0.19\pm0.01$~Hz/mm$^{2}$ and $0.35\pm 0.01$~Hz/mm$^{2}$, respectively. Both values are comfortably lower than the nEXO requirement ($<10~\text{Hz}/\text{mm}^2$). At the same over voltage setting and temperature, we measure a mean charge produced by CAs per pulse equal to $0.23\pm0.06$~PE and to $0.06\pm0.02$~PE, and an RMS of $0.51\pm0.06$~PE and $0.25\pm0.01$~PE, for FBK VUVHD3 SiPMs and HPK VUV4 MPPCs respectively. These quantities give a corresponding CAF, defined in Eq.~\ref{eq:ratio_CA}, of $0.42\pm0.07$ and $0.24\pm0.02$. The PDE of FBK VUVHD3 SiPMs and HPK VUV4 MPPCs were also characterised in the wavelength range [160-200]~nm at 163~K and 300~K. For a mean wavelength of $175$~nm and 3~V of over voltage we measured an average PDE of
$24.3\pm1.4$\% for FBK VUVHD3 SiPMs and of $20.5\pm1.1$\% for HPK VUV4 MPPCs. Both values are well above the $\ge$15\% required by nEXO.\\\indent Finally, we estimated the energy resolution that could potentially be achieved  at the 0$\nu \beta \beta$ decay $Q$-value of \ce{^{136}Xe} by the nEXO detector with the VUV sensitive SiPMs tested in this work. At an over voltage of 3~V, we estimate an energy resolution of   $0.73\pm0.02$\% for FBK VUVHD3 SiPMs and of $0.76\pm0.01$\% for HPK VUV4 MPPCs.  Overall, the devices tested in this work feature remarkable improvement compared to previously tested FBK SiPMs and HPK MPPCs and meet the nEXO requirements, making them suitable choices for the nEXO detector.\\\indent The next steps towards the conceptual design of the nEXO light detection module involve the development with HPK and FBK of photosensors with $1~\text{cm}^2$ photosensitive area. Moreover R\&D with FBK is ongoing to develop SiPMs with Through Silicon Vias (TSV) technology~\cite{Parellada2022}. This option is already available for HPK MPPCs. TSVs would remove the need for wire-bonds on the photosensors front side, simplifying assembly and avoiding wire-bonds in high electric field regions of the detector. While initial radioassays have been performed for all candidate materials in the light detection system, measurements of the radiopurity of the final assembled modules must still be performed. Finally, additional measurements are planned to determine: (i) the photosensors infrared PDE and reflectivity, which are useful to constrain the impact of the secondary photon emission in the nEXO energy resolution;  (ii) the long term stability of nEXO light detection modules in nEXO-like operating conditions~({\it i.e.}, operation in LXe under illumination from $\gamma$ calibration sources).

\section{Acknowledgments}
The authors thank Robert~E.~Vest (NIST) and Austin de St.~Croix (Queen's University) whose feedback improved the quality of this work. 
The authors gratefully acknowledge support for nEXO from NSERC, CFI, FRQNT, NRC, and the McDonald Institute (CFREF) in Canada; from the Office of Nuclear Physics within DOE’s Office of Science, and NSF in the United States; from the National Research Foundataion (NRF) of South Africa;  from IBS in Korea; from RFBR in Russia; and from CAS and NSFC in China.

\bibliographystyle{unsrt}
\bibliography{sample}

\begin{thebibliography}{10}

\bibitem{gallina2021development}
Giacomo Gallina.
\newblock {\em Development of a single vacuum ultra-violet photon-sensing
  solution for nEXO}.
\newblock PhD thesis, University of British Columbia, 2021.

\bibitem{Baudis2018}
L.~Baudis and et~al.
\newblock {Characterisation of Silicon Photomultipliers for liquid xenon
  detectors}.
\newblock {\em Journal of Instrumentation}, 13(10):P10022, 2018.

\bibitem{Baldini2018}
Meg2 Collaboration, A.~M. Baldini, and et~al.
\newblock The design of the meg2 experiment: Meg2 collaboration.
\newblock {\em European Physical Journal C}, 78, 5 2018.

\bibitem{FALCONE2021164648}
A.~Falcone and et~al.
\newblock Cryogenic sipm arrays for the dune photon detection system.
\newblock {\em Nuclear Instruments and Methods in Physics Research Section A},
  985:164648, 2021.

\bibitem{aalseth2018darkside}
DarkSide Collaboration, C.~E. Aalseth, and et~al.
\newblock Darkside-20k: A 20 tonne two-phase lar tpc for direct dark matter
  detection at lngs.
\newblock {\em The European Physical Journal Plus}, 133(3):131, 2018.

\bibitem{SiPMFBKDS20k}
F.~Acerbi and et~al.
\newblock {Cryogenic Characterization of FBK HD Near-UV Sensitive SiPMs}.
\newblock {\em IEEE Transactions on Electron Devices}, 64(2):521--526, 2017.

\bibitem{Adhikari2022}
nEXO collaboration, G.~Adhikari, and et~al.
\newblock {nEXO: neutrinoless double beta decay search beyond $10^{28}$ year
  half-life sensitivity }.
\newblock {\em Journal of Physics G: Nuclear and Particle Physics},
  49(1):015104, 2022.

\bibitem{Redshaw2007}
M.~Redshaw and et~al.
\newblock {Mass and Double-Beta-Decay $Q$-value of \ce{^{136}Xe}}.
\newblock {\em Physical Review Letters}, 98(5):053003, 2007.

\bibitem{PhysRevC.82.024603}
P.~M. McCowan and R.~C. Barber.
\newblock $q$-value for the double-$\ensuremath{\beta}$ decay of
  $^{136}\mathrm{Xe}$.
\newblock {\em Phys. Rev. C}, 82:024603, Aug 2010.

\bibitem{NEXOCollaboration2018}
nEXO Collaboration, S.~Al Kharusi, and et~al.
\newblock {nEXO Pre-Conceptual Design Report}.
\newblock \url{http://arxiv.org/abs/1805.11142}, 2018.

\bibitem{Conti2003}
E.~Conti and et~al.
\newblock Correlated fluctuations between luminescence and ionization in liquid
  xenon.
\newblock {\em Physical Review B}, 68:054201, 8 2003.

\bibitem{Abe2019}
K.~Abe andet al.
\newblock Development of low radioactivity photomultiplier tubes for the
  xmass-i detector.
\newblock {\em Nuclear Instruments and Methods in Physics Research Section A:
  Accelerators, Spectrometers, Detectors and Associated Equipment},
  922:171--176, 2019.

\bibitem{Aprile2015}
E.~Aprile and et~al.
\newblock Lowering the radioactivity of the photomultiplier tubes for the
  xenon1t dark matter experiment.
\newblock {\em The European Physical Journal C}, 75:546, 2015.

\bibitem{Neilson2009}
R.~Neilson and et~al.
\newblock Characterization of large area apds for the exo-200 detector.
\newblock {\em Nuclear Instruments and Methods in Physics Research Section A:
  Accelerators, Spectrometers, Detectors and Associated Equipment}, 608:68--75,
  9 2009.

\bibitem{FUJII2015293}
Keiko F.
\newblock High-accuracy measurement of the emission spectrum of liquid xenon in
  the vacuum ultraviolet region.
\newblock 795:293--297, 2015.

\bibitem{Ako}
nEXO collaboration, A.~Jamil, and et~al.
\newblock {VUV-Sensitive Silicon Photomultipliers for Xenon Scintillation Light
  Detection in nEXO}.
\newblock {\em IEEE Transactions on Nuclear Science}, 65(11):2823--2833, 2018.

\bibitem{Ieki2019LargeareaMW}
Keiko I. and et~al.
\newblock Large-area mppc with enhanced vuv sensitivity for liquid xenon
  scintillation detector.
\newblock {\em Nuclear Instruments and Methods in Physics Research Section A},
  2019.

\bibitem{Baudis_2021}
L.~Baudis and et~al.
\newblock Design and construction of xenoscope {\textemdash} a full-scale
  vertical demonstrator for the {DARWIN} observatory.
\newblock {\em Journal of Instrumentation}, 16(08):P08052, 2021.

\bibitem{Brown1980}
G.~N. Brown and W.~T. Ziegler.
\newblock Vapor pressure and heats of vaporization and sublimation of liquids
  and solids of interest in cryogenics below 1-atm pressure, 1980.

\bibitem{fabristhesis}
L.~Fabris.
\newblock {\em Novel readout design criteria for SiPM-based radiation
  detectors}.
\newblock PhD thesis, Universita' degli studi di Bergamo, 2015.

\bibitem{Gallina2019}
nEXO collaboration, G.~Gallina, and et~al.
\newblock {Characterization of the Hamamatsu VUV4 MPPCs for nEXO}.
\newblock {\em Nuclear Instruments and Methods in Physics Research Section A},
  940:371--379, 2019.

\bibitem{Capasso2020}
M.~Capasso and et~al.
\newblock {FBK VUV-sensitive Silicon Photomultipliers for cryogenic
  temperatures}.
\newblock {\em Nuclear Instruments and Methods in Physics Research Section A},
  982:164478, 2020.

\bibitem{Hamamatsu_doc}
Hamamatsu.
\newblock {VUV-MPPC 4th generation (VUV4)}.
\newblock
  \url{https://hamamatsu.su/files/uploads/pdf/3_mppc/s13370_vuv4-mppc_b_(1).pdf}.

\bibitem{Nakarmi2020}
nEXO collaboration, P~Nakarmi, and et~al.
\newblock {Reflectivity and PDE of VUV4 Hamamatsu SiPMs in liquid xenon}.
\newblock {\em Journal of Instrumentation}, 15(01):P01019--P01019, 2020.

\bibitem{Wagenpfeil2021}
nEXO collaboration, M.~Wagenpfeil, and et~al.
\newblock {Reflectivity of VUV-sensitive silicon photomultipliers in liquid
  Xenon}.
\newblock {\em Journal of Instrumentation}, 16(08):P08002, 2021.

\bibitem{Lv2020}
nEXO collaboration, P.~Lv, and et~al.
\newblock {Reflectance of Silicon Photomultipliers at Vacuum Ultraviolet
  Wavelengths}.
\newblock {\em IEEE Transactions on Nuclear Science}, 67(12):2501--2510, 2020.

\bibitem{nEXO_Sensitivity}
nEXO collaboration, J.~B. Albert, and et~al.
\newblock {Sensitivity and discovery potential of the proposed nEXO experiment
  to neutrinoless double-$\beta$ decay}.
\newblock {\em Physical Review C}, 97(6):065503, 2018.

\bibitem{Kotov2016}
Ivan~V. Kotov and et~al.
\newblock {Characterization and acceptance testing of fully depleted thick CCDs
  for the large synoptic survey telescope}.
\newblock In Andrew~D. Holland and James Beletic, editors, {\em High Energy,
  Optical, and Infrared Detectors for Astronomy VII}, volume 9915, page 99150V,
  2016.

\bibitem{Amp}
\url{http://www.minicircuits.com/pdfs/MAR-6SM+.pdf}.

\bibitem{Retiere2009}
F.~Reti{\`{e}}re and et~al.
\newblock {Characterization of Multi Pixel Photon Counters for T2K Near
  Detector}.
\newblock {\em Nuclear Instruments and Methods in Physics Research, Section A},
  610(1):378--380, 2009.

\bibitem{Hurkx1992}
G.A.M. Hurkx and et~al.
\newblock {A new analytical diode model including tunneling and avalanche
  breakdown}.
\newblock {\em IEEE Transactions on Electron Devices}, 39(9):2090--2098, 1992.

\bibitem{gundacker2020silicon}
S.~Gundacker and A.~Heering.
\newblock The silicon photomultiplier: fundamentals and applications of a
  modern solid-state photon detector.
\newblock {\em Physics in Medicine \& Biology}, 65(17):17TR01, 2020.

\bibitem{McLaughlin2021}
J.~B. McLaughlin and et~al.
\newblock {Characterisation of SiPM Photon Emission in the Dark}.
\newblock {\em Sensors}, 21(17):5947, 2021.

\bibitem{Boone2017}
K.~Boone and et~al.
\newblock {Delayed avalanches in Multi-Pixel Photon Counters}.
\newblock {\em Journal of Instrumentation}, 12(07):P07026--P07026, 2017.

\bibitem{Falcone2021}
A.~Falcone andet al.
\newblock Cryogenic sipm arrays for the dune photon detection system.
\newblock {\em Nuclear Instruments and Methods in Physics Research Section A},
  985:164648, 1 2021.

\bibitem{Butcher2017}
A.~Butcher and et~al.
\newblock {A method for characterizing after-pulsing and dark noise of PMTs and
  SiPMs}.
\newblock {\em Nuclear Instruments and Methods in Physics Research Section A},
  875:87--91, 2017.

\bibitem{Merzi2022}
S.~Merzi and et~al.
\newblock Nuv-hd sipms with metal-filled trenches.
\newblock {\em Conference on New Developments in Photodetection (NDIP20)},
  2022.

\bibitem{Gallina2019avalanche}
G.~Gallina and et~al.
\newblock {Characterization of SiPM Avalanche Triggering Probabilities}.
\newblock {\em IEEE Transactions on Electron Devices}, 66(10):4228--4234, 2019.

\bibitem{Zhang2022}
J.~Z. and et~al.
\newblock Scintillation detectors with silicon photomultiplier readout in a
  dilution refrigerator at temperatures down to 0.2 k.
\newblock {\em Journal of Instrumentation}, 17:P06024, 6 2022.

\bibitem{Acerbi2022}
F.~Acerbi and et~al.
\newblock Nuv and vuv sensitive silicon photomultipliers technologies optimized
  for operation at cryogenic temperatures.
\newblock {\em Nuclear Instruments and Methods in Physics Research Section A},
  page 167683, 11 2022.

\bibitem{Doc_VUV4}
K.~Yamamoto and et~al.
\newblock {New improvements to a specialized Multi-Pixel Photon Counter (MPPC)
  for neutrinoless double-beta decay and dark matter search experiments}.
\newblock ICHEP, ID:450
  \url{https://www.hamamatsu.com/sp/ssd/microsite/MPPC/ICHEP_MPPC_panel_160731.pdf},
  2016.

\bibitem{Aprile}
E.~Aprile and et~al.
\newblock Energy resolution and linearity of xenon1t in the mev energy range.
\newblock {\em The European Physical Journal C}, 80, 08 2020.

\bibitem{Pereira}
Guilherme P.
\newblock Energy resolution of the lz detector for high-energy electronic
  recoils.
\newblock {\em International Workshop on Applications of Noble Gas Xenon to
  Science and Technology" (XeSAT 2022)}, 2022.

\bibitem{nEXO:2020mec}
nEXO collaboration, T.~Stiegler, et~al.
\newblock {Event reconstruction in a liquid xenon Time Projection Chamber with
  an optically-open field cage}.
\newblock {\em Nucl. Instrum. Meth. A}, 1000:165239, 2021.

\bibitem{Neves2017}
F.~Neves and et~al.
\newblock Measurement of the absolute reflectance of polytetrafluoroethylene
  (ptfe) immersed in liquid xenon.
\newblock {\em Journal of Instrumentation}, 12:P01017--P01017, 2017.

\bibitem{Doke:2002oab}
Tadayoshi Doke and et~al.
\newblock {Absolute Scintillation Yields in Liquid Argon and Xenon for Various
  Particles}.
\newblock {\em Jap. J. Appl. Phys.}, 41:1538--1545, 2002.

\bibitem{Martin1971}
M.~Martin.
\newblock Exciton self‐trapping in rare‐gas crystals.
\newblock {\em The Journal of Chemical Physics}, 54:3289--3299, 4 1971.

\bibitem{PhysRevB.20.3486}
Shinzou K. and et~al.
\newblock Dynamical behavior of free electrons in the recombination process in
  liquid argon, krypton, and xenon.
\newblock {\em Phys. Rev. B}, 20:3486--3496, Oct 1979.

\bibitem{PhysRevB.27.5279}
Akira H. and et~al.
\newblock Effect of ionization density on the time dependence of luminescence
  from liquid argon and xenon.
\newblock {\em Phys. Rev. B}, 27:5279--5285, May 1983.

\bibitem{wasterdalethesis}
S.S. Westerdale.
\newblock {\em A Study of Nuclear Recoil Backgrounds in Dark Matter Detectors}.
\newblock PhD thesis, Princeton University, 2016.

\bibitem{Collaboration2019}
G.~Anton and et~al.
\newblock {Measurement of the scintillation and ionization response of liquid
  xenon at MeV energies in the EXO-200 experiment}.
\newblock {\em Physical Review C}, 101(6):065501, 2020.

\bibitem{Szydagis2011}
M.~Szydagis and et~al.
\newblock Nest: a comprehensive model for scintillation yield in liquid xenon.
\newblock {\em Journal of Instrumentation}, 6:P10002--P10002, 10 2011.

\bibitem{Baudis2021_2}
Baudis L. and et~al.
\newblock A measurement of the mean electronic excitation energy of liquid
  xenon.
\newblock {\em The European Physical Journal C}, 81:1060, 2021.

\bibitem{RevModPhys.82.2053}
E.~Aprile and T.~Doke.
\newblock Liquid xenon detectors for particle physics and astrophysics.
\newblock {\em Rev. Mod. Phys.}, 82:2053--2097, 2010.

\bibitem{Seguinot1995}
J.~Seguinot and et~al.
\newblock {Liquid xenon scintillation: photon yield and Fano factor
  measurements}.
\newblock {\em Nuclear Instruments and Methods in Physics Research Section A},
  354(2-3):280--287, 1995.

\bibitem{Boulay2022}
M.~G. Boulay and et~al.
\newblock Sipm cross-talk in liquid argon detectors.
\newblock \url{http://arxiv.org/abs/2201.01632}, 2022.

\bibitem{Parellada2022}
L.~Parellada-Monreal and et~al.
\newblock Silicon photomultipliers technologies for 3d integration.
\newblock {\em Conference on New Developments in Photodetection (NDIP20)},
  2022.

\end{thebibliography}

\end{document}